\documentclass[prc,twocolumn,nofootinbib,superscriptaddress,aps]{revtex4-1}

\usepackage{amsmath}
\usepackage{amssymb}
\usepackage{graphicx}
\usepackage{dcolumn,dsfont}
\usepackage{bm,braket}
\usepackage{hyperref}
\usepackage{isotope}
\usepackage{multirow}
\usepackage{tabularx}
\usepackage{comment}
\usepackage{url}
\usepackage{relsize}
\usepackage{xcolor}

\newcommand{\kf}{k_{\rm F}}
\newcommand{\MeV}{\,\mathrm{MeV}}

\newcommand{\fmiq}{\,\mathrm{fm}^{-3}}

\newcommand{\km}{\,\mathrm{km}}
\newcommand{\Msun}{\,M_\odot}
\newcommand{\Gth}{\Gamma_\mathrm{th}}
\newcommand{\Ye}{Y_\mathrm{e}}
\newcommand{\ns}{n_0}
\newcommand{\dns}{\eta}
\newcommand{\n}{_\mathrm{n}}
\newcommand{\p}{_\mathrm{p}}

\newcommand{\esym}{E_\mathrm{sym}} 
\newcommand{\SE}{S} 
\newcommand{\SEv}{E_\mathrm{sym}} 

\begin{document}

\title{New equations of state constrained by nuclear physics,\\
observations, and QCD calculations of high-density nuclear matter}

\author{S. Huth}
\email[Email:~]{shuth@theorie.ikp.physik.tu-darmstadt.de}
\affiliation{Technische Universit{\"a}t Darmstadt, Department of Physics, 64289 Darmstadt, Germany}
\affiliation{ExtreMe Matter Institute EMMI, GSI Helmholtzzentrum f\"ur Schwerionenforschung GmbH, 64291 Darmstadt, Germany}
\author{C. Wellenhofer}
\email[Email:~]{wellenhofer@theorie.ikp.physik.tu-darmstadt.de}
\affiliation{Technische Universit{\"a}t Darmstadt, Department of Physics, 64289 Darmstadt, Germany}
\affiliation{ExtreMe Matter Institute EMMI, GSI Helmholtzzentrum f\"ur Schwerionenforschung GmbH, 64291 Darmstadt, Germany}
\author{A. Schwenk}
\email[Email:~]{schwenk@physik.tu-darmstadt.de}
\affiliation{Technische Universit{\"a}t Darmstadt, Department of Physics, 64289 Darmstadt, Germany}
\affiliation{ExtreMe Matter Institute EMMI, GSI Helmholtzzentrum f\"ur Schwerionenforschung GmbH, 64291 Darmstadt, Germany}
\affiliation{Max-Planck-Institut f\"ur Kernphysik, Saupfercheckweg 1, 69117 Heidelberg, Germany}

\begin{abstract}
We present new equations of state for applications in core-collapse supernova and neutron star merger simulations. We start by introducing an effective mass parametrization that is fit to recent microscopic calculations up to twice saturation density. This is important to capture the predicted thermal effects, which have been shown to determine the proto-neutron star contraction in supernova simulations. The parameter range of the energy-density functional underlying the equation of state is constrained by chiral effective field theory results at nuclear densities as well as by functional renormalization group computations at high densities based on QCD. We further implement observational constraints from measurements of heavy neutron stars, the gravitational wave signal of GW170817, and from the recent NICER results. Finally, we study the resulting allowed ranges for the equation of state and for properties of neutron stars, including the predicted ranges for the neutron star radius and maximum mass. 
\end{abstract}

\maketitle

\section{Introduction}

The equation of state (EOS) of dense matter is of central interest in nuclear physics and astrophysics. In the multi-messenger era, gravitational wave observations of neutron star mergers such as GW170817~\cite{LIGO18NSradii,LIGO19update} as well as measurements of neutron star radii from NASA's NICER mission~\cite{Rile19NICER,Mill19NICER,Raai19NICEREOS} are providing novel constraints on the EOS of neutron star matter. 

In recent years, many efforts have been undertaken to construct EOS parametrizations (see, e.g., Refs.~\cite{Hebe13ApJ,Rrap16constmfmod,Lim18TidalDef,Tews18sos,Grei19sos}) based on microscopic calculations of pure neutron matter (PNM)~\cite{Hebe10nmatt,Tews13N3LO,Lynn16QMC3N,Dris17MCshort}. However, such calculations are only available up to $1$--$2$ times nuclear saturation density $\ns$, such that extrapolations to high densities using, e.g., piecewise polytropes~\cite{Hebe13ApJ} or a speed of sound parametrization~\cite{Tews18sos,Grei19sos} have to be applied. High-density constraints from neutron star observations, in particular precise mass measurements of heavy neutron stars~\cite{Anto13PSRM201,Crom19massiveNS}, reduce the parameter space of possible extensions significantly. 

Compared to cold isolated neutron stars, core-collapse supernovae (CCSNe) and neutron star mergers (NSM) probe a much wider range of the ($n,x,T$) space of the EOS, where $n$ is the baryon density, $x$ the proton fraction, and $T$ the temperature. The EOSs commonly used in numerical CCSN and NSM simulations are based either on phenomenological Skyrme energy-density functionals such as the Lattimer-Swesty EOS~\cite{Latt91LSEOS} or relativistic mean-field models like the Shen EOS~\cite{Shen98eos}. More recently, new EOSs for astrophysical simulations based on these two approaches have been constructed~\cite{GShen11SHTeos,Hemp12EOSccsn,Stei13EOSccsn,Schn17LSEOS}. 

However, these phenomenological EOSs are often not consistent with microscopic calculations at zero temperature~\cite{Hebe13ApJ,Krue13N3LOlong,Hebe10nmatt,Tews13N3LO,Lynn16QMC3N,Dris17MCshort} and nuclear thermodynamics~\cite{Well14nmtherm,Well15therm,Carb19thermalEOS,Kell20finiteT}. Thermal effects are well characterized by the nucleon effective mass $m^*_t(n,x)$~\cite{Carb19thermalEOS,Kell20finiteT}, where $t=\text{n},\text{p}$ for neutrons and protons. The effective mass is a crucial quantity in CCSN simulations that governs the proto-neutron star contraction~\cite{Yasi20EOSeffects,Schn19EOSeffects}. Furthermore, recently a first step towards systematic computations of the EOS at intermediate to high densities was achieved by Leonhardt {\it et al.}~\cite{Leon19fRGeos} who calculated symmetric nuclear matter (SNM) at densities $3\lesssim n/\ns\lesssim 10$ by using functional renormalization group (fRG) methods based on QCD. Constructing new EOS parametrizations that are consistent with state-of-the-art constraints from nuclear physics, astrophysical observations, and high-density QCD calculations will enable important progress in nuclear astrophysics.

In this work, we develop a novel EOS functional that incorporates recent microscopic results for the nucleon effective mass~\cite{Carb19thermalEOS}. The parameters of the EOS functional are then fit to theoretical calculations at low and high densities as well as observational constraints from mass measurements, GW170817, and NICER. The systematics of different parameter choices is analyzed in detail, and based on our EOS functional we derive comprehensive uncertainty bands for the EOS and for neutron star properties. Our work provides the groundwork for new EOSs of hot and dense matter for CCSN and NSM simulations.

This paper is organized as follows. In Sec.~\ref{ch2} we review current constraints on the EOS by nuclear physics, observations, and high-density QCD calculations. The effective mass parametrization together with the EOS functional and its characteristics are the topic of Sec.~\ref{ch3}. In Sec.~\ref{ch4} we discuss the different fit procedures used to constrain the parameters of the EOS functional and study in detail the impact of variations of the parameters. Based on this, in Sec.~\ref{ch5} we examine our predictions for various neutron star properties and thermal effects. Finally, in Sec.~\ref{ch6} we conclude with a short summary and outlook.

\section{Overview of equation of state constraints}
\label{ch2}

Here, we briefly summarize presently available constraints on the EOS of dense nuclear matter. First, in Sec.~\ref{sec21} we examine constraints from nuclear physics on the properties of neutron-rich matter at densities up to $1$--$2$ times nuclear saturation density. Then, in Sec.~\ref{sec22} we discuss high-density constraints inferred from recent neutron star observations. Finally, we discuss the results of a recent fRG study of SNM at higher densities in Sec.~\ref{sec23}. 

\subsection{Constraints from nuclear physics}\label{sec21}

In this section we summarize constraints on the EOS from nuclear theory and experiment. In particular, we discuss constraints on various characteristic parameters of the EOS  around saturation density $n_0$: the binding energy $B$, the incompressibility $K$, the symmetry energy coefficient $\esym$, and the slope parameter $L$. (The effective mass is discussed in Sec.~\ref{sec31}.) Further, we discuss constraints from theoretical calculations on the EOS of PNM.

\subsubsection{Approximate EOS parametrizations}\label{sec21a}

The bulk of dense matter in CCSN and NSM consists of nucleons and electrons (as well as a small fraction of muons). For isolated neutron stars the temperature is negligible with respect to nuclear energy scales. Thus, the neutron star EOS is predominantly determined by the ground-state energy of strongly-interacting nucleonic matter $E(n,\beta)$. In neutron stars, the isospin asymmetry $\beta=\left(n\n-n\p\right)/n$ is for a given nucleon number density $n=n\n+n\p$ fixed by the conditions of beta equilibrium and charge neutrality (see Sec.~\ref{sec51} for details). A very useful approximative parametrization of the dependence of $E(n,\beta)$ on $\beta$ is given by
\begin{equation}
\dfrac{E}{A}(n,\beta)\approx\dfrac{E}{A}(n,0)+\SE(n)\beta^2 \,.
\label{eq:expandE}
\end{equation}
Here, $E(n,0)$ is the energy of SNM, $E(n,1)$ corresponds to PNM, and $\SE(n)$ is called the symmetry energy. It has been validated in various microscopic nuclear many-body calculations~\cite{Bomb91anm,Carb13tensoresym,Dris14asymmat,Dris16asym,Well16DivAsym,Somasundaram:2020chb} that Eq.~\eqref{eq:expandE} provides a very good approximation of the exact $\beta$ dependence over the entire range $\beta\in[0,1]$ for densities up to $n\lesssim (1$--$2)\ns$.\footnote{The accuracy of Eq.~\eqref{eq:expandE} is, however, expected to decrease with increasing density~\cite{Well16DivAsym,Somasundaram:2020chb}.} The symmetry energy $\SE(n)$ may be obtained as the quadratic coefficient in the expansion of $E(n,\beta)$ in $\beta$.\footnote{No odd powers of $\beta$ occur in the expansion if charge-symmetry breaking effects are neglected. Notably, beyond second-order logarithmic terms $\sim \beta^{2n\geqslant 4}\ln|\beta|$ may appear at zero temperature~\cite{Kais15quartic,Well16DivAsym}.} From Eq.~\eqref{eq:expandE} it follows that the quadratic coefficient is approximately equal to the difference in energy of PNM to SNM:
\begin{equation}
\frac{1}{2}\dfrac{\partial^2 E/A}{\partial \beta^2}\Big{\vert}_{\beta=0}\approx \dfrac{E}{A}(n,1)-\dfrac{E}{A}(n,0)~.
\label{eq:symmE}
\end{equation}
Throughout this work, we assume a quadratic dependence of the interaction part of $E(n,\beta)$ on $\beta$, so for the interaction part the two sides of Eq.~\eqref{eq:symmE} are identified.
The kinetic part is modeled as a noninteracting gas of neutrons and protons with density- and isospin-dependent effective masses. Within this approximation, the isospin-asymmetry dependence of the kinetic part is treated exactly. This approach is backed up by the microscopic calculations of isospin-asymmetric nuclear matter of Refs.~\cite{Dris14asymmat,Well15therm,Well16DivAsym}. See Sec.~\ref{ch3} for details on the form of our EOS functionals. For the symmetry energy, we then choose the definition 
\begin{equation}\label{eq:symmEdef}
\SE(n)\equiv \dfrac{E}{A}(n,1)-\dfrac{E}{A}(n,0).
\end{equation}
[See, e.g., Ref.~\cite{Well16DivAsym} for an analysis of the difference between Eq.~\eqref{eq:symmEdef} and the second-order Taylor coefficient in the $\beta$ expansion.]

\begin{figure*}
    \centering
    \includegraphics[width=0.95\columnwidth]{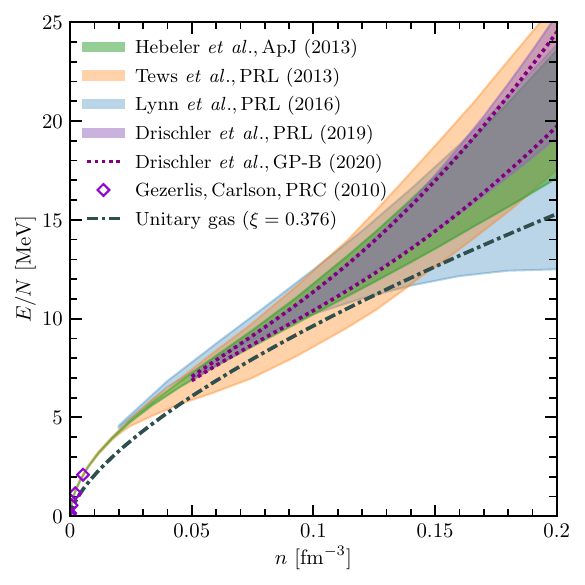}
    \hspace{5mm}
    \includegraphics[width=0.95\columnwidth]{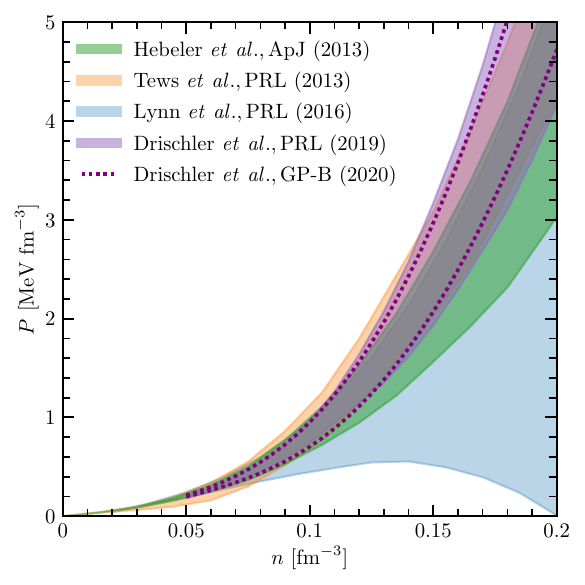}
    \caption{Energy per particle $E/N$ (left panel) and pressure $P$ (right panel) of PNM as a function of density $n$ from various many-body calculations with chiral EFT interactions~\cite{Hebe13ApJ,Tews13N3LO,Lynn16QMC3N,Dris17MCshort,Dris20GPB}; see text for details. In the left panel, we also show the low-density quantum Monte Carlo results by Gezerlis and Carlson~\cite{Geze09neutmat} as well as the conjectured lower bound given by the energy per particle of a unitary Fermi gas of neutrons~\cite{Tews17NMatter}.}
    \label{fig:energy_constraints}
\end{figure*}

Another useful parametrization is obtained by expanding the energy per particle in terms of the relative density difference to saturation density $\dns=(n-\ns)/3\ns$:
\begin{equation}
\dfrac{E}{A}(n,\beta)\approx -B + \dfrac{1}{2}K\dns^2+\left( \esym + L\dns \right)\beta^2~.
\label{eq:expandnE}
\end{equation}
Here, $B=-E/A|_{n=\ns,\beta=0}$ is the binding energy of SNM. Since the pressure $P=n^2 \partial_n E/A$ vanishes in SNM at $\ns$, the leading density dependence in the expansion of $E(n,0)$ in terms of $\dns$ corresponds to the incompressibility $K$. The expansion of the symmetry energy $\SE(n)$ introduces the symmetry energy coefficient $\esym=\SE(\ns)$ and the slope parameter $L$. The latter is proportional to the pressure of PNM at saturation density, i.e.,
\begin{equation}
L=3\ns\dfrac{\partial \SE}{\partial n}\Big{\vert}_{\ns}=\dfrac{3}{\ns}P\left( \ns,\beta=1 \right)~.\label{eq:PL}
\end{equation}
The incompressibility at saturation density is proportional to the second derivative of the energy per particle with respect to density:
\begin{equation}
K=9\dfrac{\partial P}{\partial n}\Big{\vert}_{\ns, \beta=0}=9\ns^2\dfrac{\partial^2 E/A}{\partial n^2}\Big{\vert}_{\ns,\beta=0}~.
\end{equation}
The four parameters $B$, $K$, $\esym$ and $L$ provide a thorough characterization of the neutron star EOS at densities around saturation density. 

For the saturation density $\ns$ and energy $B$, we use $\ns=0.164(7)$ and $B=15.86(57)$, which have been extracted from fits to nuclear masses and include an additional systematic uncertainty (see Ref.~\cite{Dris17MCshort}). Modern microscopic nuclear-matter calculations are consistent with these results~\cite{Hebe11fits,Dris16asym,Dris17MCshort}, but have larger uncertainties. Further, nuclear theory constrains the nuclear incompressibility as $K=215(40)$~\cite{Hebe11fits,Dris16asym,Dris17MCshort}. Many efforts have been devoted to determining the symmetry parameter $\esym$ and $L$, which play a crucial role in neutron star structure and dynamics. In particular, microscopic calculations of the EOS of PNM put tight constraints on $\esym$ and $L$. We discuss these PNM and symmetry energy constraints in the two paragraphs below. 

The temperature dependence of the EOS is key for CCSN and NSM applications. A very useful characteristic of the temperature dependence is given by the so-called thermal index $\Gamma_\text{th}$, which is defined as~\cite{Baus10thermal}
\begin{align}\label{GammathDef}
\Gth(T,n,\beta)=1+\frac{P(T,n,\beta)-P(0,n,\beta)}{\varepsilon(T,n,\beta)-\varepsilon(0,n,\beta)},
\end{align} 
with the internal energy density $\varepsilon=E/V$. The quantity $\Gth$ is a measure of thermal contributions to the equation of state. For a noninteracting nucleon gas with density-dependent effective mass $m^*(n)$ it is given by~\cite{Const15thermal} 
\begin{align} \label{GammathM*}
\Gth(n)=\frac{5}{3}-\frac{n}{m^*(n)}\frac{\partial m^*(n)}{\partial n}.
\end{align}
It has been verified in microscopic nuclear-matter calculations that the form given by Eq.~\eqref{GammathM*} provides a very precise approximation of $\Gth(T,n,\beta)$ for $\beta\in\{0,1\}$, $n\lesssim 2\ns$, and $T\lesssim~30\MeV$~\cite{Carb19thermalEOS,Kell20finiteT}. A crucial novelty in our EOS functionals is therefore the accurate implementation of the effective masses of neutrons and protons $m_{\text{n},\text{p}}^*(n,\beta)$. We discuss this and the available constraints on $m_{\text{n},\text{p}}^*(n,\beta)$ in Sec.~\ref{sec31}.
Our results for the thermal index are then examined in Sec.~\ref{sec52}.

\subsubsection{Neutron matter constraints}\label{sec21b}

The modern approach to the description of the strong interaction at nuclear energy scales is based on chiral effective field theory (EFT) and renormalization group (RG) methods~\cite{Epel09RMP,Bogn10PPNP,Mach11PR}. From general EFT convergence restrictions as well as regulator and many-body convergence considerations, the viability of this approach is restricted to densities $n\lesssim 2\ns$. The theoretical uncertainties in current implementations of chiral interactions in a given many-body framework arise from the interplay of finite-regulator artifacts, many-body and EFT truncation errors, and parameter-fitting ambiguities.

Because nuclear forces are weaker in PNM, the theoretical uncertainties are under better control there compared to SNM. In Fig.~\ref{fig:energy_constraints} we compare the results for the energy per particle and pressure of PNM obtained from several recent nuclear many-body calculations with chiral EFT interactions. The results by Hebeler {\it et al.}~\cite{Hebe13ApJ}, Tews {\it et al.}~\cite{Tews13N3LO}, and Drischler {\it et al.}~\cite{Dris17MCshort,Dris20GPB} are based on many-body perturbation theory, while the results by Lynn {\it et al.}~\cite{Lynn16QMC3N} were obtained from auxiliary-field diffusion Monte Carlo computations using local chiral interactions. In each case, the results include uncertainty estimates, shown as bands in Fig.~\ref{fig:energy_constraints}. These are based on EFT truncation errors and different regulators in Refs.~\cite{Lynn16QMC3N,Dris17MCshort,Dris20GPB}, while they are mainly due to uncertainties in the low-energy couplings that enter three-nucleon forces in Refs.~\cite{Hebe13ApJ,Tews13N3LO}. The uncertainty band of Drischler {\it et al.}~PRL (2019)~\cite{Dris17MCshort} is based on simple EFT truncation errors. The results of Drischler {\it et al.}~GP-B (2020)~\cite{Dris20GPB} are constructed from the same calculations (from Ref.~\cite{Dris17MCshort}) but based on a Bayesian uncertainty analysis using Gaussian processes, which leads to a very similar band for the combined GP-B (450) and (500) results. One sees that while overall the results from these calculations are in good agreement, the uncertainties become sizable for densities $n\gtrsim\ns$.

At densities near and above saturation density the uncertainties associated with the effective description of the nuclear interactions dominate over many-body truncation effects. At low densities $n\ll\ns$, the nuclear interactions are less intricate, but here the many-body accuracy may be inflicted by the sensitivity to large-scattering length physics. Still, as shown in Fig.~\ref{fig:energy_constraints}, the various chiral EFT-based many-body calculations discussed above are in reasonable agreement with the low-density results from precise quantum Monte Carlo computations by Gezerlis and Carlson~\cite{Geze09neutmat}.

Finally, in Fig.~\ref{fig:energy_constraints} we also show the energy per particle of a unitary Fermi gas of neutrons $E_\text{UG}(n)=\xi E_\text{FG}(n)$, where $E_\text{FG}(n)$ is the free neutron gas energy and the Bertsch parameter is $\xi\approx 0.376$~\cite{Ku563}. In Ref.~\cite{Tews17NMatter}, it was argued that $E_\text{UG}(n)$ can be used as a lower bound for the PNM energy. As seen in Fig.~\ref{fig:energy_constraints}, the unitary gas bound reduces the uncertainties in the results by Tews {\it et al.}~\cite{Tews13N3LO} and Lynn {\it et al.}~\cite{Lynn16QMC3N}, while the ones by Hebeler {\it et al.}~\cite{Hebe13ApJ} and Drischler {\it et al.}~\cite{Dris17MCshort,Dris20GPB} are unaffected.

\subsubsection{Symmetry energy constraints}\label{sec21c}

Equation~\eqref{eq:expandnE} provides the basis for a benchmark analysis of related aspects of nuclear physics and astrophysics~\cite{Tsan12esymm,Latt12esymm}. The symmetry energy coefficient $\SEv$ and slope parameter $L$ are crucial for a variety of phenomena in nuclear physics and astrophysics, ranging from nuclear masses~\cite{Kort10edf}, neutron skins and the dipole polarizability~\cite{Tami11dipole,Roca15alphaD,Hage16NatPhys,Birk17dipole,Kauf2068Ni}, to heavy-ion collisions~\cite{Tsan09Lconstr}, and core-collapse supernovae~\cite{Yasi20EOSeffects,Schn17LSEOS}. In particular, the neutron star radius scales with the pressure of PNM at saturation density~\cite{Latt12esymm,Latt07EOSconst}, i.e., with the $L$ parameter [see Eq.~\eqref{eq:PL}]. 

\begin{figure}
    \centering
    \includegraphics[width=0.95\columnwidth]{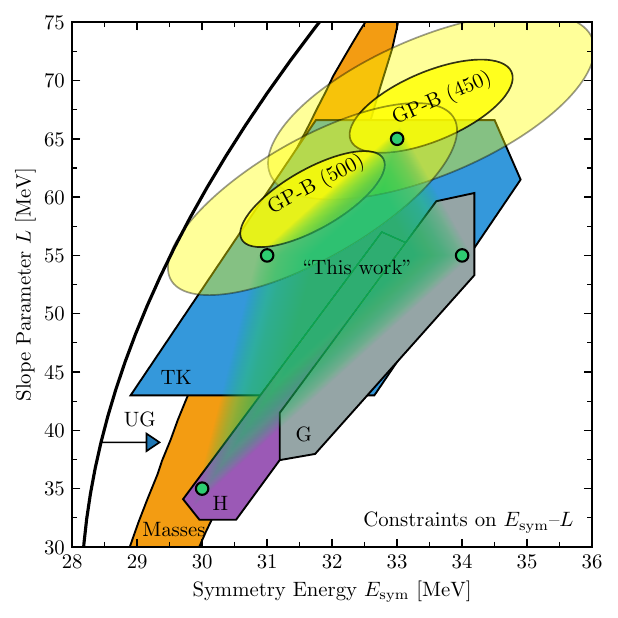}
    \caption{Theoretical constraints on the symmetry energy $\SEv$ and the slope parameter $L$ from Refs.~\cite{Hebe13ApJ,Gand12nm,Tews13N3LO,Dris20GPB}, see text for details. Also shown are constraints extracted from fits to nuclear masses (orange band)~\cite{Kort10edf} and the conjectured unitary gas (UG) bound~\cite{Tews17NMatter}. The corners (green dots) of the green shaded area (``This work'') correspond to the four representative $(\SEv,L)$ pairs adopted in this work. The figure is adapted from Refs.~\cite{Latt12esymm,Dris20GPB} using the Jupyter notebook provided in Refs.~\cite{Buge20git,Dris20GPB}.}\label{fig:EsymL}
\end{figure}

Many experimental and theoretical efforts have been undertaken to constrain the symmetry energy. In Fig.~\ref{fig:EsymL}, we show results for the correlation between $\SEv$ and $L$ obtained from the microscopic PNM calculations by Hebeler {\it et al}.~(H)~\cite{Hebe13ApJ}, Tews {\it et al}.~(TK)~\cite{Tews13N3LO}, and Drischler {\it et al}.~[GP-B (450), GP-B (500)]~\cite{Dris20GPB} discussed in Sec.~\ref{sec21b} above. In addition, we show results from auxiliary-field diffusion Monte Carlo calculations by Gandolfi {\it et al}.~(G)~\cite{Gand12nm}. The uncertainties in the H, TK, and G results were obtained by using various (chiral) two- and three-nucleon interactions. The TK results involve the largest uncertainties, which stems in part from larger variations of the low-energy couplings in three-nucleon forces. In the GP-B case, we show results obtained from chiral potentials with two different cutoffs, GP-B (450) and (500), where in each case the uncertainties were obtained from a Bayesian analysis using Gaussian processes of the fixed-cutoff EFT systematics. Note that in Fig.~\ref{fig:energy_constraints} the two GP-B bands are combined in one single band.

The region of $(\SEv,L)$ values spanned by these theoretical results overlaps with the constraints extracted from various experiments; see, e.g., Refs.~\cite{Latt12esymm,Latt14symenerg,Dris20GPB}. As an example, in Fig.~\ref{fig:EsymL} we show the constraint extracted from nuclear masses~\cite{Kort10edf}. Also shown is the conjectured unitary gas (UG) bound~\cite{Tews17NMatter}. The theoretical constraints (H,K,TK,GP-B) are all consistent with the UG constraint.

For the EOS constructed in this work we choose four representative $(\SEv,L)$ pairs that lie within the combined theoretical constraints (H,K,TK,GP-B):
\begin{align} \label{eq:EsymLpairs}
(\SEv,L)/\MeV \in \big\{(30,35), (31,55), (33,65), (34,55) \big\}.
\end{align}
These four pairs are shown as green dots in Fig.~\ref{fig:EsymL} and the (green shaded) region spanned by them is labeled ``This work''. Note that we exclude very large symmetry energies and slope parameters to avoid having too many EOSs exceed the PNM uncertainty band shown in Fig.~\ref{fig:energy_constraints}.\footnote{Even with $(\SEv,L)$ values inside the GP-B region our EOS functional can still lead to PNM properties, which are incompatible with the theoretical PNM uncertainty band. This is because of higher-order terms in the density behavior.} A detailed study of the EOS and neutron star properties associated with our four $(\SEv,L)$ pairs is provided in Sec.~\ref{sec41}.

\subsection{Constraints from neutron star observations}\label{sec22}

Neutron star observations play a crucial role in constraining the dense-matter EOS. In particular, mass measurements of two-solar-mass neutron stars~\cite{Demo10ns,Anto13PSRM201,Crom19massiveNS} have narrowed the uncertainties in the neutron star mass-radius relation considerably. To support neutron stars of such mass the EOS cannot be too soft, which challenges neutron star models that include substantial portions of exotic condensates or deconfined quark matter. The present lower bound for the maximal mass $M_\mathrm{max}$ is given by the mass of the heaviest observed neutron stars: PSR J0740+6620 with a mass of $M=2.14^{+0.20}_{-0.18}\Msun$~\cite{Crom19massiveNS} at the $2\sigma$ level measured using relativistic Shapiro delay. This is in line with the radio-timing observation of the pulsar J0348+0432 with $M=2.01\pm 0.04\Msun$~\cite{Anto13PSRM201}. In our work, we use the averaged lower bound of $M=1.965\Msun$ as a constraint for the lower bound of the maximal mass.

Since the observation of the first NSM GW170817~\cite{LIGO18NSradii,LIGO19update} by LIGO/Virgo together with the observation of the corresponding kilonova AT2017gfo and the short gamma-ray burst GRB170817A~\cite{Abbo17GW170817}, there have been many efforts to infer an upper bound on the maximum neutron star mass $M_\mathrm{max}$ from the remnant behavior. The suggested limits are generally in the range $M_\mathrm{max}\lesssim 2.3$--$2.4\Msun$~\cite{Marg17Mmax,Shib17GWmodel, Ruiz18Mmax, Rezz18Mmax, Shib19Mmax, Ai20Mmax, Abbo20modelcompare,Shao20Mmax}, which would rule out overly stiff EOS, in addition to the soft EOSs ruled out by the two-solar-mass constraint. 

Even with a relatively narrow range on the maximal mass, the radius of a typical neutron star with $M=1.4\Msun$ is uncertain, with a typical conservative range $10\lesssim R/\text{km}\lesssim 14$; see, e.g., Refs.~\cite{Hebe13ApJ,Tews18sos,Grei19sos,Capa20NatAst}. Recently, a major step toward precise radius measurements was made by the NICER collaboration~\cite{Rile19NICER,Mill19NICER}, which simultaneously determined the radius and mass of PSR J0030+0451 via x-ray pulse-profile modeling. 

\begin{figure}
    \centering
    \includegraphics[width=0.95\columnwidth]{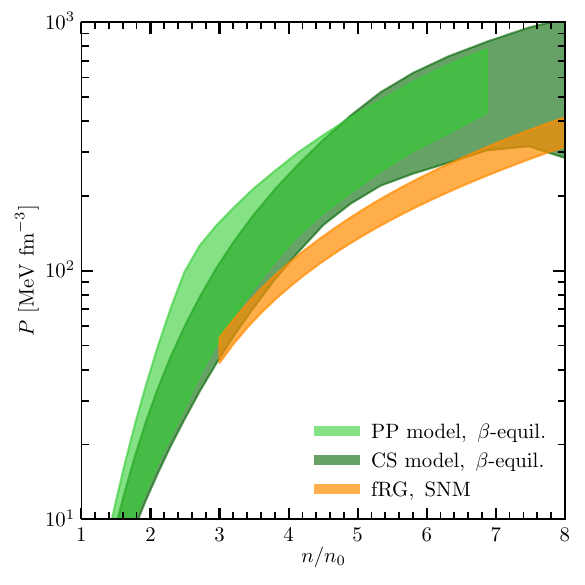}
    \caption{Constraints on the pressure of neutron star matter as a function of density $n/n_0$ (green bands) from a joint analysis~\cite{Raai19GWNICER} of the $2.14\Msun$ mass constraint, GW170817, and the NICER results, obtained using two different EOS models: piecewise polytropes (PP) and speed of sound model (CS), see Sec.~\ref{sec22} for details. The bands are for the $95\%$ credible regions. Also shown are the results for the pressure of SNM (orange band) from the fRG study of Ref.~\cite{Leon19fRGeos}; see Sec.~\ref{sec23}.}
    \label{fig:pressure_constraints}
\end{figure}

Implications of this measurement on the EOS have been studied by Raaijmakers {\it et al.}~\cite{Raai19NICEREOS} by applying two parametrizations for the neutron star EOS (in $\beta$-equilibrium): a piecewise polytropic (PP) model~\cite{Hebe13ApJ} and a speed of sound (CS) parametrization~\cite{Grei19sos}. Raaijmakers {\it et al.}~\cite{Raai19GWNICER} performed a joint analysis of these models to infer implications on the EOS from the NICER measurement, GW170817, and the $2.14\Msun$ mass measurement. Their results for the pressure as a function of density are shown in Fig.~\ref{fig:pressure_constraints} (green bands). While the PP and CS bands are consistent with each other, the PP model allows stiffer EOSs for densities $n\lesssim 4\ns$ and in general smaller maximal densities. Consequently, the $M\mbox{-}R$ relation of the CS model involves somewhat smaller radii compared to the PP model. In our work, we use the combined PP and CS bands by Raaijmakers {\it et al.}~\cite{Raai19GWNICER} as a constraint for our EOS parametrization. 

\subsection{Theoretical calculations at high densities}\label{sec23}

The ultra-high-density regime ($n\gtrsim50\ns$) of the EOS corresponds to deconfined quark matter. Perturbative QCD provides the expansion of the EOS about the high-density limit~\cite{PhysRevLett.121.202701}. This expansion can be used to construct astrophysical EOSs from interpolating between the chiral EFT band for the EOS at nuclear densities and the perturbative QCD region~\cite{Kurk14nsQCD,Anna18TidalDef}. As discussed in Sec.~\ref{sec22}, in the present paper we incorporate high-density constraints from neutron star observations explicitly. Moreover, we base the high-density extrapolation of the EOS of SNM on a recent fRG calculation at more relevant densities. Therefore, in our case the perturbative QCD expansion does not provide significant additional constraints on the dense-matter EOS.

At present no reliable and accurate method exists for computing the properties of strongly interacting matter at densities $n\gtrsim2\ns$ (apart from the perturbative QCD limit). However, a notable step towards systematic high-density calculations was made by Leonhardt \textit{et al.}~in Ref.~\cite{Leon19fRGeos}. Starting from the QCD action, they use the fRG to derive a low-energy quantum effective action with effective four-quark interactions and diquark degrees of freedom. The uncertainties in the results for the (zero-temperature) EOS of SNM from this approach have been estimated in terms of their RG scale dependence. Other sources of error are, e.g., due to neglected quark flavors and higher-order interaction effects.  

The fRG results of Leonhardt \textit{et al.}~\cite{Leon19fRGeos} for the pressure of SNM are shown in Fig.~\ref{fig:pressure_constraints} (orange band). They span a band from $n=3\ns$ to $n=10\ns$. The band lies mostly below the observational neutron star matter constraints from Raaijmakers \textit{et al.}~\cite{Raai19GWNICER}, with small overlaps for densities near $n=3\ns$ and near $n=8\ns$. Note that the fRG band is significantly smaller than the neutron star matter bands of Raaijmakers \textit{et al.}. In this paper, we will see that neutron star constraints, in particular the maximum mass constraint $M_\text{max}\geqslant 1.965\ M_\odot$, tend to favor a pressure of SNM that lies somewhat above the fRG band either near $n=3\ns$ or near $n=8\ns$; see Secs.~\ref{sec32},~\ref{sec43}, and~\ref{ch5}. Nevertheless, we find several EOS that are consistent with neutron star observations and for which the pressure of SNM lies within the fRG band for $n\gtrsim 5\ns$.

\section{New Equation of state Functional}\label{ch3}

We now come to the construction of a new EOS functional that takes into account the constraints from nuclear physics, neutron star observations and high-density QCD calculations summarized in Sec.~\ref{ch2}. First, in Sec.~\ref{sec31} we examine recent microscopic calculations of the neutron and proton effective mass $m_{\text{n},\text{p}}^*(n,\beta)$ in SNM ($\beta=0$) and PNM ($\beta=1$), and introduce a convenient parametrization of $m_{\text{n},\text{p}}^*(n,\beta)$ to be implemented in our EOS functionals. The construction of the EOS functional is the subject of Sec.~\ref{sec32}.

\subsection{Temperature dependence and\\ nucleon effective mass}\label{sec31}

Recently, Carbone and Schwenk~\cite{Carb19thermalEOS} computed the finite-temperature EOS of PNM and SNM from chiral EFT interactions using the self-consistent Green's function method. In addition, they also calculated the effective masses $m_{\text{n},\text{p}}^*(n,\beta=0,1)$. Based on these results they showed that the thermal index $\Gth$ obtained from the pressure and the energy density, see Eq.~\eqref{GammathDef}, can be accurately parametrized in terms of the effective mass, via the form given by Eq.~\eqref{GammathM*}. Recent microscopic neutron-matter calculations in many-body perturbation theory have confirmed this result~\cite{Kell20finiteT}. Therefore, a reliable implementation of the effective masses of neutrons and protons $m_{\text{n},\text{p}}^*(n,\beta)$ is crucial to capture thermal effects in astrophysical applications.

To this end, we introduce an effective mass parametrization that fits the results for $m_{\text{n},\text{p}}^*(n,\beta=0,1)$ at densities $n\lesssim 2\ns$ from Ref.~\cite{Carb19thermalEOS} based on the N$^3$LO NN potential from Ref.~\cite{Ente03EMN3LO} and N$^2$LO 3N interactions constructed in Ref.~\cite{Klos17triton}. The behavior of $m_{\text{n},\text{p}}^*(n,\beta)$ at higher densities is uncertain. We explore different scenarios in this regime. Our effective mass parametrization as a function of density is given by
\begin{align}
\dfrac{m^*_t}{m} &= 1 + \left( \alpha_1 n_t + \beta_1 n_{-t} + \alpha_2 n_{t}^{4/3} + \beta_2 n_{-t}^{4/3} + \alpha_3 n_{t}^{5/3} 
\right. \nonumber \\ & \quad \left.
+ \beta_3 n_{-t}^{5/3}  \right) \dfrac{1}{1+e^{5n}} 
\nonumber \\ &
+ \left( \epsilon_t \dfrac{n_t}{n} + \epsilon_{-t} \dfrac{n_{-t}}{n} -1 \right)\dfrac{1-e^{-10n}}{1+e^{-5(n-n_\mathrm{off})}}~,
\label{eq:effm_para}
\end{align}
where the nucleon with opposite isospin is denoted by $-t$. The six parameters $\alpha_i, \beta_i$ with $i\in (1,3)$ are fit to the SNM and PNM results of Ref.~\cite{Carb19thermalEOS}. The factor $1/(1+e^{5n})$ (sigmoid function) has the effect that the fitted part goes to zero with increasing density. The high-density behavior of $m_{\text{n},\text{p}}^*(n,\beta)$ is then fixed by the parameters $\epsilon_t$ and $\epsilon_{-t}$ as well as by the offset $n_\mathrm{off}$ of the (modified) logistic function $(1-e^{-10n})/(1+e^{-5(n-n_\mathrm{off})})$. For instance, for $\beta=1$ the high-density limit of the neutron effective mass is given by $\epsilon\n$, and for $\beta=0$ the nucleon effective mass approaches the value $(\epsilon\n+\epsilon\p)/2$. Note that this implies that the high-density limit of the thermal index is $\Gth\rightarrow 5/3$; see Eq.~\eqref{GammathM*}. The correct ultrarelativistic limit is $\Gth\rightarrow 4/3$, but this matters for the nucleonic part of the EOS only for densities far above those relevant for neutron stars.

\begin{figure}
    \centering
    \includegraphics[width=0.95\columnwidth]{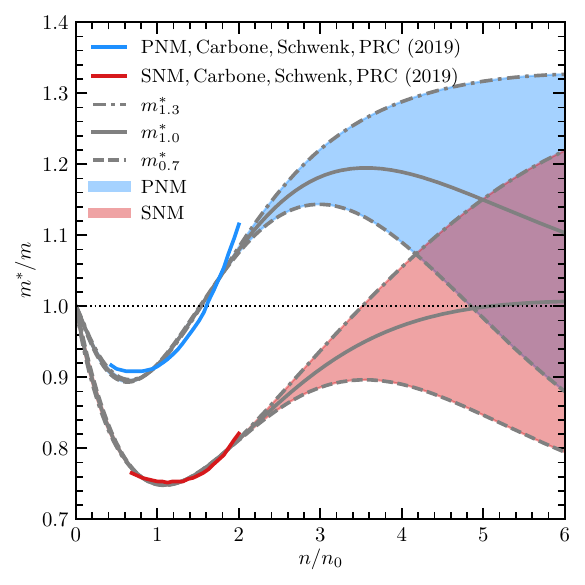}
    \caption{Effective mass $m^*/m$ as a function of density $n/n_0$ for PNM and SNM. The blue (PNM) and red (SNM) solid lines up to $n/n_0 = 2$ show the results of Carbone and Schwenk~\cite{Carb19thermalEOS}. The gray lines (connected by colored bands) correspond to the three representative effective mass parametrizations employed in this work. The associated high-density limits are given by $m^*/m \rightarrow 0.7$ (dashed lines), $m^*/m \rightarrow 1.0$ (solid lines), and $m^*/m \rightarrow 1.3$ (dash-dotted lines).}
    \label{fig:effmass_functional}
\end{figure}

In Fig.~\ref{fig:effmass_functional}, we show the results for the nucleon effective mass in PNM and SNM from Ref.~\cite{Carb19thermalEOS} and three representative effective mass parametrizations based on Eq.~\eqref{eq:effm_para}. [In the PNM case we show the results for the neutron effective mass $m_\text{n}^*(n,\beta=1)$.] At low densities, the effective mass is a decreasing function of density
(see also the recent auxiliary field diffusion Monte Carlo computations~\cite{Bura19effmassPNM}), but starting at around nuclear saturation density it increases with density for both PNM and SNM mainly due to 3N forces. The effective mass in PNM is larger than the one for SNM (see also Ref.~\cite{Hebe09enerfunc}) and for densities $n\gtrsim 1.5\ns$ it exceeds the bare nucleon mass. 

From Eq.~\eqref{GammathM*} it follows that the thermal index is $\Gth<5/3$ in the density region where the effective mass increases with density. Depending on the form of the increase at higher densities it may even be that $\Gth<1$ at high densities. This would imply a negative thermal expansion coefficient~\cite{Well14nmtherm}, with a negative thermal pressure contribution, see Eq.~\eqref{GammathDef}, so that the pressure at finite temperature would be smaller than the pressure at $T=0$. While such a feature is not unphysical in general, it would still be somewhat peculiar to have $\Gth<1$ in nuclear matter. We therefore restrict the high-density extrapolations of the effective mass to a form that ensures that $\Gth>1$. 

In this work, we set $n_\mathrm{off}=0.7\fmiq$ and restrict ourselves to cases where $\epsilon_t=\epsilon_{-t}=\epsilon$ in Eq.~\eqref{eq:effm_para} such that the effective mass has the same high-density limit in SNM and PNM. We employ three representative values of the high-density limit, i.e., $\epsilon\in\{0.7,1.0,1.3\}$, denoted by $m^*_{0.7}$, $m^*_{1.0}$, and $m^*_{1.3}$ in Fig.~\ref{fig:effmass_functional}. As seen in Fig.~\ref{fig:effmass_functional}, these three scenarios span a reasonably wide range for the behavior of the effective mass at high densities. Note also that the fit below twice saturation density is unaffected by the high-density behavior (with our parametrization this holds true even for more extreme values of $\epsilon$). Consequently, our effective mass scenarios only affect thermal properties at very high densities.

Regarding isospin-asymmetric nuclear matter, our effective mass parametrization ensures that the neutron effective mass $m_\text{n}^*(n,\beta)$ increases with $\beta$ and satisfies $m_\text{n}^*(n,\beta)>m_\text{p}^*(n,\beta)$, in agreement with theoretical results~\cite{Sjoe76Landaumass,Dale05effmass,Li16effmassBHF,Li18effm,Well16DivAsym}. In contrast to $m_\text{n}^*(n,\beta)$, which is constrained by fits to microscopic calculations for both PNM and SNM, in our approach the $\beta$ dependence (at finite $n$) of the proton effective mass $m_\text{p}^*(n,\beta)$ is an outcome of the fit of $m_\text{p}^*(n,0)=m_\text{n}^*(n,0)$ to SNM [after $m_\text{n}^*(n,1)$ has been fit to PNM]. We found that the $\beta$ dependence of both $m_\text{n}^*$ and $m_\text{p}^*$ is largest at $n\approx 3-4\ns$ (the high-density limit $\epsilon$ is $\beta$ independent). Moreover, our parametrization leads to a $\beta$ dependence of $m_\text{p}^*(n,\beta)$ that is decreased compared to that of $m_\text{n}^*(n,\beta)$, which is consistent with the results from Refs.~\cite{Dale05effmass,Li16effmassBHF}. For $m^*_{0.7}$, the proton effective mass decreases with $\beta$ at low densities $n\lesssim \ns$ and increases for $n\gtrsim \ns$. In the $m^*_{1.0}$ and $m^*_{1.3}$ case the proton effective mass decreases with $\beta$ at all densities, with the decrease being significantly more pronounced for $m^*_{1.3}$. Here, the behavior for $m^*_{1.0}$ and $m^*_{1.3}$ is more in line with nuclear theory results~\cite{Dale05effmass,Li16effmassBHF}. Future work may involve the construction of improved EOS functionals that incorporate additional theoretical constraints on the $\beta$ dependence of $m_\text{p}^*(n,\beta)$.

\subsection{Equation of state functional}\label{sec32}

We now introduce the new EOS functional that forms the basis for the investigations carried out in the remainder of this paper. The microscopic results for the relation between the effective mass $m_t^*(n,\beta)$ and the thermal index $\Gth(n,\beta)$ discussed in Secs.~\ref{sec21} and~\ref{sec31} make clear that a reasonable approach to the temperature dependence of an effective EOS functional is to use a $T$-dependent kinetic term with density-dependent effective mass and a $T$-independent interaction part. This is also supported by the microscopic nuclear-matter calculations of Refs.~\cite{Well14nmtherm,Well15therm,Well16DivAsym,Carb19thermalEOS,Kell20finiteT}, where it was found that the $T$ dependence of the interaction contribution in many-body perturbation theory is small compared to the one of the noninteracting contribution.

In the usual (Skyrme) energy density functionals, the interaction part is modeled as a finite polynomial in fractional powers of density; see, e.g., Refs.~\cite{Lim17NScrust,Schn17LSEOS}. By construction, the high-density behavior of a polynomial EOS ansatz involves a highly fine-tuned balance between different density powers. In certain cases, i.e., for some judicious choices of the density powers, a reasonable high-density extrapolation can result from fits to microscopic calculations at nuclear densities~\cite{Lim17NScrust,Lim18TidalDef}. However, for the systematic construction of EOS functionals constrained by nuclear physics, neutron star observations, and high-density QCD calculations, a polynomial ansatz can clearly encounter difficulties.

We therefore choose a form of the interaction part that ameliorates this fine tuning. For the internal energy density as a function of density $n=n\n+n\p$, proton fraction $x=n\p/n$ and temperature $T$ we use the following form:
\begin{equation}
\begin{aligned}
\dfrac{E}{V} &(n,x,T) = \sum_t \dfrac{\tau_t(n,x,T)}{2m^*_t(n,x)} -xn\Delta
\\&+ \sum_i \left[\dfrac{a_i}{d_a+n^{(\delta_i-2)/3}}
+\dfrac{4b_ix(1-x)}{d_b+n^{(\delta_i-2)/3}}\right]n^{1+\delta_i/3}~.
\label{eq:EVfunctional}
\end{aligned}
\end{equation}
Here, the second term gives the rest mass contribution (modulo the neutron mass energy), with $\Delta$ the neutron--proton mass difference. The first term corresponds to the kinetic part of the internal energy density; it is modeled as a noninteracting gas of neutrons and protons with effective masses $m^*_{\text{n},\text{p}}(n,x)$ given by Eq.~\eqref{fig:effmass_functional}. That is, the term $\tau_t$ is given by\footnote{We use the nonrelativistic quasiparticle dispersion relation for all densities. The high-density behavior of our EOSs is fit to observational and fRG constraints, 
so only thermal effects at very high densities are affected by this approximation, 
which is, however, a minor effect in comparison to the effective-mass uncertainties in that regime.}
\begin{align}
\tau_t(n,x,T) &= \frac{1}{2\pi^2}\int^\infty_0 dp\, p^4 
\nonumber \\ &\quad \times
\frac{1}{1+\exp\big[\frac{1}{T}\big(\frac{p^2}{2m^*_t(n,x)}-\tilde\mu_t(n,x,T)\big)\big]}\,,
\end{align}
where the auxiliary chemical potential $\tilde\mu_t(n,x,T)$ is defined via
\begin{equation}\label{eq:tildemu}
n_t = \frac{1}{2\pi^2}\int^\infty_0 dp\, p^2 \frac{1}{1+\exp\big[\frac{1}{T}\big(\frac{p^2}{2m^*_t}-\tilde\mu_t\big)\big]}\,.
\end{equation}
The $T\rightarrow 0$ limit of the kinetic part is given by $\tau_t(n,x,0)=(3\pi^2 n_t)^{5/3}/(5\pi^2)$. The role of the auxiliary chemical potential is similar to the one in many-body perturbation theory at finite temperature~\cite{Well18StatQuas}. The true chemical potential is obtained from the thermodynamic potential corresponding to the variables $(n,x,T)$, i.e., the free energy. Since the interaction part and the effective masses are $T$ independent, the free energy is obtained by substituting  the kinetic part of $E/V$ with the free energy density of a (nonrelativistic) noninteracting gas of neutrons and protons with effective masses $m^*_{\text{n},\text{p}}(n,x)$.

The third term in Eq.~\eqref{eq:EVfunctional} is the interaction part. The crucial feature of the interaction part is that it is based on rational functions instead of density monomials. While the parameters $a_i$ and $b_i$ with $i\in (1,4)$ are fit to low- and high-density results as specified below, the density exponents $\delta_i$ as well as $d_a$ and $d_b$ are not fit parameters but set to specific values. We choose two different sets for $\delta_i$:
\begin{align} \label{eq:deltasets}
\delta_{\kf}&=(3,4,5,6),\quad\quad
\delta_{n}   =(3,6,9,12).
\end{align}
For the choice $\delta_{\kf}$, the density exponents in the numerators of the interaction part are $(1, 4/3, 5/3, 2)$, corresponding to integer powers of the Fermi momentum $\kf$ at zero temperature. The choice $\delta_{n}$ corresponds to integer powers of $n$ in the numerators. The density dependence of the denominators is chosen such that in the high-density limit the interaction part becomes proportional to $n^{5/3}$.\footnote{The density dependence in the ultrarelativistic limit is $\sim n^{4/3}$, but this matters only for densities far above those relevant for neutron stars; see Sec.~\ref{sec31}.} The purpose of the denominators is to mitigate the fine-tuning between the different parts of the interaction term such that the EOS functional is stable under variations of the fit input. For a given choice of $\delta_{i}$, the fit performance of the EOS functional is controlled by the two offset parameters $d_a$ and $d_b$. We set $d_a=d_b=d$ and use for $d$ the following values:
\begin{align} \label{eq:dsets}
d_{\kf}&\in\{1,3,5,7\},\quad\quad
d_{n}   \in\{0.2,0.4,0.6,0.8\}.
\end{align}
These choices provide a reasonably wide range of different density behaviors, as examined in detail below. The smaller values of $d$ for $\delta_{n}$ are mandated by the large density exponents, i.e., the suppression of higher density powers must set in earlier there. We note that removing the restriction $d_a=d_b$ has no notable impact on our results.

We fix the eight parameters $a_{1,2,3,4}$ and $b_{1,2,3,4}$ by matching to the following input:
\begin{itemize}
\item the energy per particle of PNM at $n=0.05\fmiq$, determined by the QMC result from Ref.~\cite{Geze09neutmat} as $E/N(0.05\fmiq) = 2.1\MeV$,
\item the nuclear matter properties $(\ns,B,K,\esym,L)$,
\item the pressure of PNM at $n=1.28\fmiq\approx 8\ns$,
\item the pressure of SNM at $n=1.28\fmiq\approx 8\ns$.
\end{itemize}
Here, the six nuclear-density inputs (first two items) are varied according to their uncertainties, as examined in Sec.~\ref{sec21}. The high-density input for the pressure of PNM and SNM is taken such that the resulting EOS is consistent with constraints from neutron star observations [Sec.~\ref{sec22}] and the pressure of SNM is in reasonable agreement with the fRG results [Sec.~\ref{sec23}] (we allow a $10\%$ deviation from the fRG band to account for further fRG uncertainties).

\begin{figure}
    \centering
    \includegraphics[width=0.95\columnwidth]{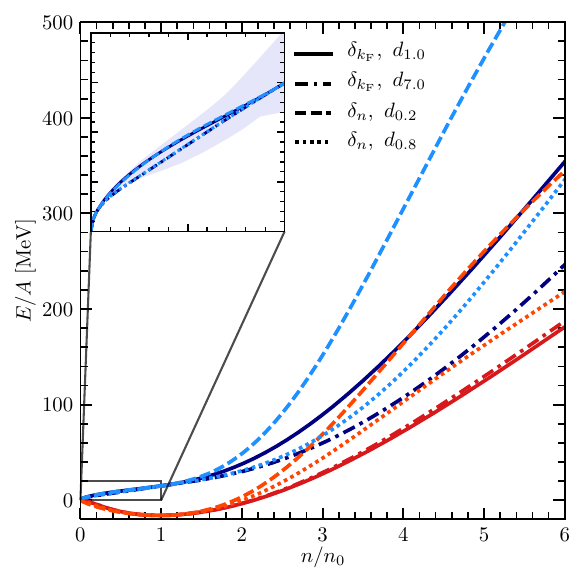}
    \caption{Results for the energy per particle of PNM (blue) and SNM (red) as a function of density $n/n_0$ obtained from the two $\delta_i$ sets with the minimal and maximal value of the corresponding $d$. We use the same set of low- and high-density fit points for all depicted EOS. The light blue band in the inset corresponds to the combined chiral EFT results from Fig.~\ref{fig:energy_constraints}.}
    \label{fig:energy_functional}
\end{figure}

\begin{figure}
    \centering
    \includegraphics[width=0.95\columnwidth]{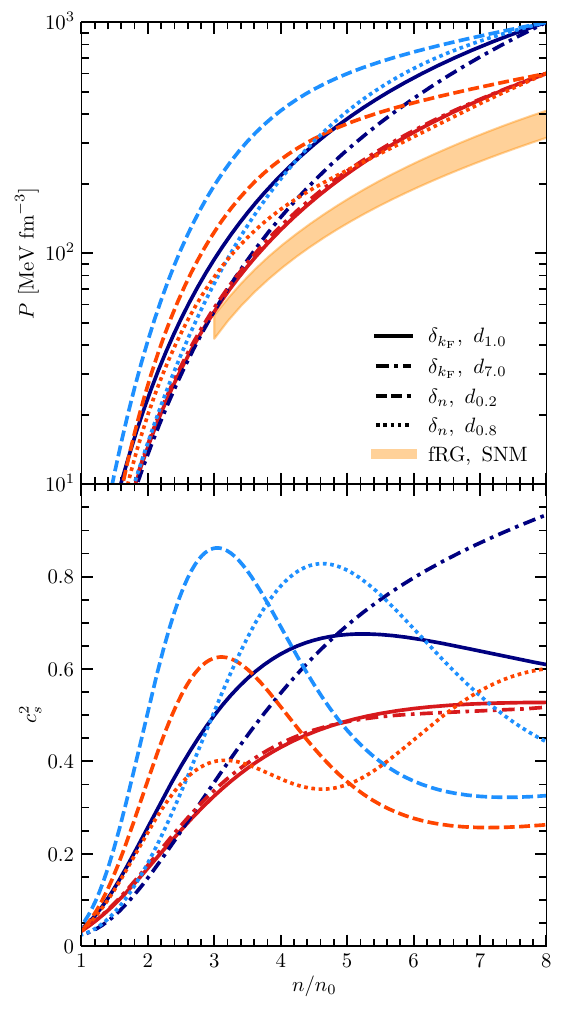}
    \caption{Same as Fig.~\ref{fig:energy_functional} but here we show the pressure $P$ (upper panel) and speed of sound $c_s^2$ (lower panel) of PNM (blue) and SNM (red). Note the pressure fit points at $8 n_0$. The orange band corresponds to the fRG results of Ref.~\cite{Leon19fRGeos}. }
    \label{fig:pressure_cs2_functional}
\end{figure}

The results for the energy per particle $E/A$ of PNM and SNM obtained for one particular input set are shown in Fig.~\ref{fig:energy_functional}. Specifically, here we set the nuclear matter properties to $(B,K,\esym,L)=(15.29,255,30,35)\MeV$ and $\ns=0.157\fmiq$, the pressure at $n=1.28\fmiq$ for SNM to $600\MeV\fmiq$ and for PNM to $1000\MeV\fmiq$, and use the effective mass scenario $m^*_{1.0}$. Figure \ref{fig:pressure_cs2_functional} shows the corresponding results for the pressure $P$ and the square of the speed of sound $c_s^2$ (in units where $c=1$), which is given by the derivative of the pressure with respect to energy density. Variations of the input are investigated in Sec.~\ref{ch4}. For each $\delta_i$ set [Eq.~\eqref{eq:deltasets}], we show the results for the smallest and the largest $d$ in the corresponding set of possible $d$ values [Eq.~\eqref{eq:dsets}]. One sees that the nuclear-density input has the effect that for densities up to roughly twice saturation density the functional is not very sensitive to the values of $\delta_i$ and $d$. At intermediate densities $2\ns\lesssim n<8\ns$, different choices of $\delta_i$ and $d$ result in the following systematics:
\begin{itemize}
\item $\delta_{\kf},d=1.0$ (solid lines): We find a soft EOS for SNM indicated by the rather small values of the square of the speed of sound, which only barely exceeds $c_s^2 \approx 0.5$. In the case of PNM, the EOS is stiffer to support a $2\Msun$ neutron star and $c_s^2$ shows a broad peak around $5\ns$.
\item $\delta_{\kf},d=7.0$ (dash-dotted lines): An enhancement of the $d$ parameter results in no notable changes for SNM. In contrast, increasing $d$ softens the EOS of PNM for $n\lesssim 5\ns$, while at high densities it is significantly stiffer and $c_s^2$ almost reaches the speed of light at $8\ns$. Nevertheless, the energy per particle as well as the pressure of PNM is smaller in the given density range.
\item $\delta_{n},d=0.2$ (dashed lines): Compared to the choice $\delta_{\kf}$, the larger density exponents in $\delta_{n}$ lead to a rapid increase of the energy per particle and the pressure of both SNM and PNM at comparatively low densities. As the density increases the EOS becomes softer again to match the pressure fit point at $8\ns$. As a consequence, we find a pronounced peak for the speed of sound due to the stiffness of the EOS. 
\item $\delta_{n},d=0.8$ (dotted lines): Again, a larger $d$ parameter softens the EOS. In comparison to $\delta_{\kf}$, the sensitivity of the functional with respect to $d$ is more pronounced for PNM as well as for SNM. In this specific case, the speed of sound has a second maximum at high densities; this feature depends on the chosen input values and is not present in most EOS.
\end{itemize}

These characteristics of the EOS functional are quite robust throughout the input parameter space. In line with this, our new approach (rational functions instead of density monomials) ensures that the fit parameters $(a_i,b_i)$ remain of reasonable size (i.e., there is no ``unnatural'' fine-tuning) under comprehensive variations of the low- and high-density input. More precisely, for the 16,128 input sets considered in Sec.~\ref{sec51}, those that are consistent with the imposed constraints from nuclear physics and observations with the density exponents $\delta_n$ have absolute values of the dimensionless parameters of at most 2.6, with mean values between 0.2 and 1.1. For $\delta_{\kf}$ the parameters are in general larger due to the smaller density exponents in the numerators of the interaction terms, i.e., $\delta_{\kf}$ involves more tuning than $\delta_n$. The mean values of the fit parameters for $\delta_{\kf}$ lie in the range 1.4 to 15.5. For the combined ($\delta_{\kf}$ and $\delta_{n}$) parameter space, $80\%$ of the constrained EOS have $a_i$ and $b_i$ in a range from $-12$ (lower bound $b_2$) to 20 (upper bound $b_3$), with five out of the eight fit parameters spanning only a range at most from $-3$ to 3. Overall, as discussed in Secs.~\ref{ch3} and \ref{ch4}, our approach allows to generate (via variations of the low- and high-density input) a broadly populated range of EOSs that reflects well the uncertainties from nuclear physics, observations, and high-density QCD calculations. 

\section{Equation of state variations}\label{ch4}

With the energy density functional in place, we perform variations of the input choices to span a range of EOS that covers the uncertainties of the constraints discussed in Sec.~\ref{ch2}. First, we discuss the variations of nuclear matter properties in Sec.~\ref{sec41}. This involves the four representative $(\esym,L)$ pairs shown in Fig.~\ref{fig:EsymL} as well as three choices for the saturation properties of SNM. In Sec.~\ref{sec42}, we then analyze the behavior of the EOS parametrization for the three effective mass scenarios introduced in Sec.~\ref{sec31}. This is followed by high-density variations of the pressure for PNM and SNM in Sec.~\ref{sec43}. These variations are performed for each set of expansion coefficients $\delta_i$ and each of the corresponding $d$ values given by Eqs.~\eqref{eq:deltasets} and~\eqref{eq:dsets}, respectively. For each type of variation, we show the corresponding influence on the energy per particle, pressure, and speed of sound for PNM and SNM with an associated figure.

\subsection{Variations of nuclear matter properties}\label{sec41}

To cover the uncertainties of the energy of PNM from many-body calculations based on chiral EFT as elaborated in Secs.~\ref{sec21b} and~\ref{sec21c}, four combinations of the symmetry energy $\esym$ and the slope parameter $L$ were identified, namely,
\begin{align} \label{eq:EsymLpairs2}
(\SEv,L)/\MeV \in \big\{(30,35), (31,55), (33,65), (34,55) \big\}.
\end{align}
These values (green dots in Fig.~\ref{fig:EsymL}) cover a reasonable range of the combined theoretical results for the $\esym$-$L$ correlation; see Fig.~\ref{fig:EsymL}. In fact, we have investigated a whole grid of $(\esym,L)$ pairs that encompasses and exceeds the green-shaded region in Fig.~\ref{fig:EsymL}: the grid ranges are $28$--$36\MeV$ for the symmetry energy and $30$--$75\MeV$ for the slope parameter (with step sizes 1 and $5\MeV$). We have examined the EOS and neutron star properties obtained from each $(\esym,L)$ pair in this grid for each $(\delta_i,d)$ choice, each effective mass scenario, and each of the different $(\ns,B,K)$ values and high-density input specified in Sec.~\ref{sec43}. For every $(\esym,L)$ pair we then counted the number of EOSs, which fulfill the constraints from nuclear physics and neutron star observations discussed in Secs.~\ref{sec21} and \ref{sec22}. The nuclear-density inputs are discussed further below; see Sec.~\ref{sec43} for details on the high-density input range considered. Note that the high-density fRG results for SNM from Sec.~\ref{sec23} are not enforced as a strict constraint.

\begin{figure}
    \centering
    \includegraphics[width=0.95\columnwidth]{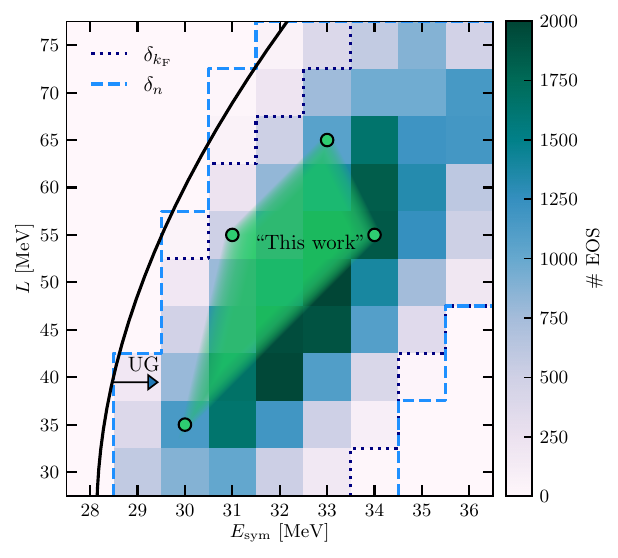}
    \caption{Grid of $(\esym,L)$ pairs used to determine the four representative pairs given by Eq.~\eqref{eq:EsymLpairs2}. The black line (labeled ``UG'') corresponds to the constraint on $(\esym,L)$ obtained from the unitary gas boundary on the PNM energy~\cite{Tews17NMatter}; see Sec.~\ref{sec21c}. The color coding gives the number of EOS that fulfill the theoretical and observational constraints discussed in Secs.~\ref{ch2}; see the text for details.}
        \label{fig:Esym_L_grid}
\end{figure}

\begin{figure*}
    \centering
    \includegraphics[width=1.7\columnwidth]{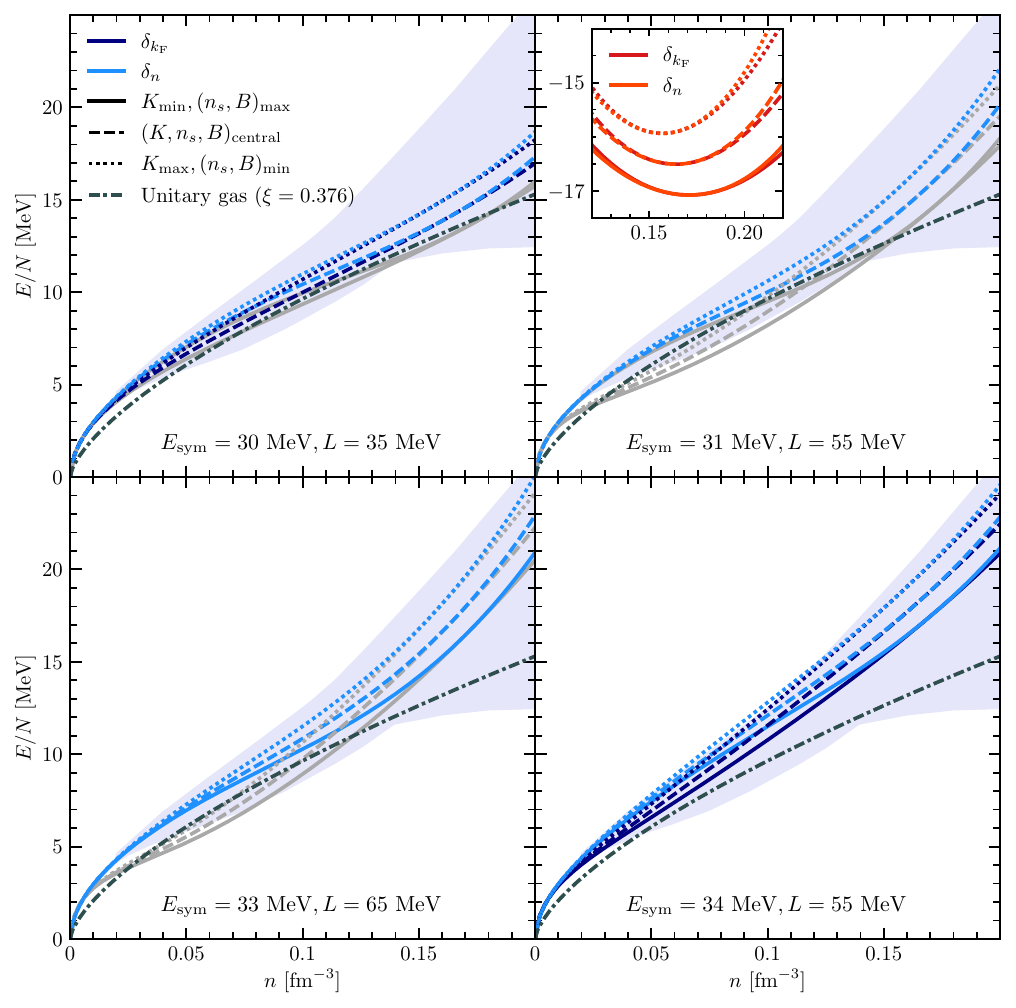}
    \caption{Results for the energy per particle of PNM at nuclear densities. The four panels correspond to the four representative $(\SEv,L)$ pairs given by Eq.~\eqref{eq:EsymLpairs2}. In each panel we show the results obtained from the three $(\ns,B,K)$ triples given by Eqs.~\eqref{eq:nsBKtriples1}--\eqref{eq:nsBKtriples3}. The light/dark blue lines are for the $\delta_n/\delta_{\kf}$ functional. All depicted EOSs employ the smallest $d$ available for the respective choice of $\delta_i$; see Eq.~\eqref{eq:dsets}, the effective mass scenario $m^*_{1.0}$, and the same high-density fits for the pressure (see Fig.~\ref{fig:pressure_cs2_nucmatt_variations}, top panel). In the inset of the second panel, we show the corresponding SNM results. The light blue band in each panel corresponds to the combined chiral EFT results from Fig.~\ref{fig:energy_constraints}. The anthracite dash-dotted line in each panel is the conjectured lower bound (unitary gas with $\xi=0.376$) on the PNM energy. All EOSs with energies that violate this bound are excluded and shown as gray lines.}
    \label{fig:energy_nucmatt_variations}
\end{figure*}

The results from this study are analyzed in Fig.~\ref{fig:Esym_L_grid} where one sees that larger slope parameters become disfavored the smaller the symmetry energy is. This feature, which is more pronounced for $\delta_{\kf}$ (see the boundary in Fig.~\ref{fig:Esym_L_grid}), is reflected also in the microscopic constraints on the $\esym$-$L$ correlation, see Fig.~\ref{fig:EsymL}. Our four choices for $(\esym,L)$ are based on the combined results from Figs.~\ref{fig:EsymL} and~\ref{fig:Esym_L_grid}, 
and on the observation that they lead to EOSs that cover a broad range of neutron star properties.

For the saturation properties of SNM, we use the empirical saturation point $\ns=0.164(7)\fmiq$ and $B=15.86(57)\MeV$ (see Ref.~\cite{Dris17MCshort}) together with the constraint on the incompressibility $K=215(40)\MeV$ determined from microscopic nuclear-matter calculations~\cite{Hebe11fits,Dris16asym,Dris17MCshort}. The uncertainties in $(\ns,B,K)$ are covered by three combinations. The triple $(K,\ns,B)_\mathrm{central}$ uses the central values. The two other triples combine the minimal (maximal) values of $\ns$ and $B$  with the largest (smallest) incompressibility: $(K_\mathrm{max},(\ns,B)_\mathrm{min})$ and $(K_\mathrm{min},(\ns,B)_\mathrm{max})$. 
Overall:
\begin{align} \label{eq:nsBKtriples1}
(K_\mathrm{min},(\ns,B)_\mathrm{max})= (175,0.171,16.43),\\ \label{eq:nsBKtriples2}
(K,\ns,B)_\mathrm{central}= (215,0.164,15.86),\\ \label{eq:nsBKtriples3}
(K_\mathrm{max},(\ns,B)_\mathrm{min})= (255,0.158,15.29),
\end{align}
in units $\MeV$, $\fmiq$, and $\MeV$, respectively. These combinations have a physical motivation: First, they follow the Coester-band correlation between $\ns$ and $B$ values~\cite{Dris17MCshort}. Second, if SNM saturates at small densities and energies, then one expects that the incompressibility increases, and vice versa.

The four $(\esym,L)$ pairs and three $(\ns,B,K)$ triples amount to twelve possible combinations of nuclear matter properties for each choice of $(\delta_i,d)$. The corresponding results for the PNM energy at nuclear densities are examined in Fig.~\ref{fig:energy_nucmatt_variations} where each of the four panels is for one of the four $(\esym,L)$ pairs. One sees that the chosen variations of the nuclear matter properties provide a thorough representation of the PNM uncertainty band from chiral EFT. The depicted EOS are for one particular choice of $(\delta_i,d)$, $m^*_t(n,x)$ and the high-density input, as specified in the caption of Fig.~\ref{fig:energy_nucmatt_variations}. Variations of these properties broaden the covered area further. We use the unitary gas bound to rule out some of the EOSs (gray lines), in particular among those are $(K_\mathrm{min},(\ns,B)_\mathrm{max})$, and $\delta_{\kf}$. The EOS with intermediate symmetry energies and large slope parameters $(\esym,L)/\MeV=(31,55)$ and $(33,65)$ are most affected by this (conjectured) lower bound.

\begin{figure} 
    \centering
    \includegraphics[width=0.95\columnwidth]{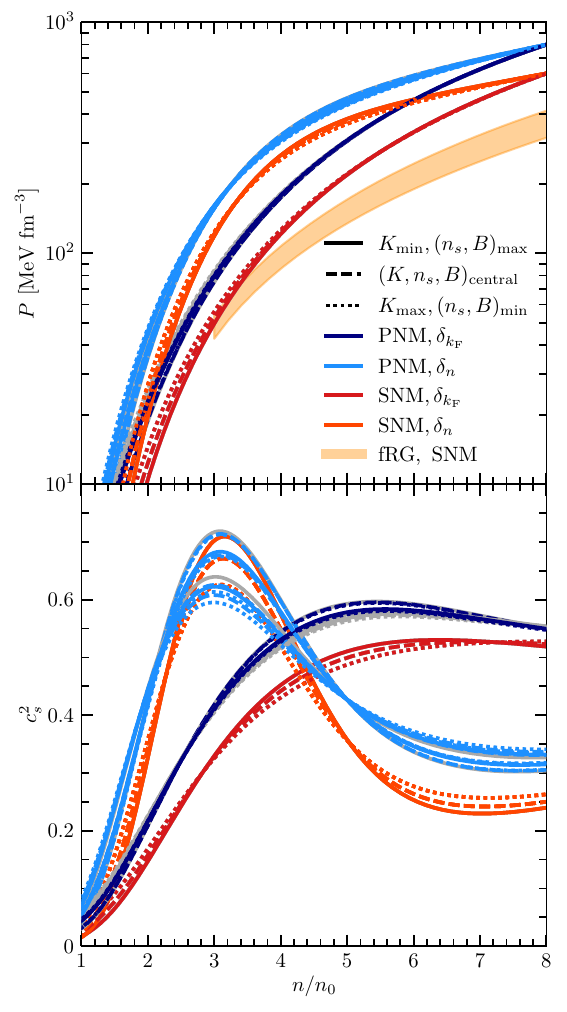}
    \caption{High-density analog of Fig.~\ref{fig:energy_nucmatt_variations} for the pressure (upper panel) and the speed of sound (lower panel) of PNM (blue) and SNM (red) as a function of density. The orange band corresponds to the fRG SNM results from Ref.~\cite{Leon19fRGeos}.}
    \label{fig:pressure_cs2_nucmatt_variations}
\end{figure}

Finally, in Fig.~\ref{fig:pressure_cs2_nucmatt_variations}, we show the analog of Fig.~\ref{fig:energy_nucmatt_variations} for the pressure and the speed of sound of PNM and SNM. One sees that the nuclear matter properties have only a relatively small impact on the high-density behavior, which for a given high-density fit input is predominantly determined by the choice of $\delta_{i}$ and $d$. 

\subsection{Effective mass variation}\label{sec42}

\begin{figure} 
    \centering
    \includegraphics[width=0.95\columnwidth]{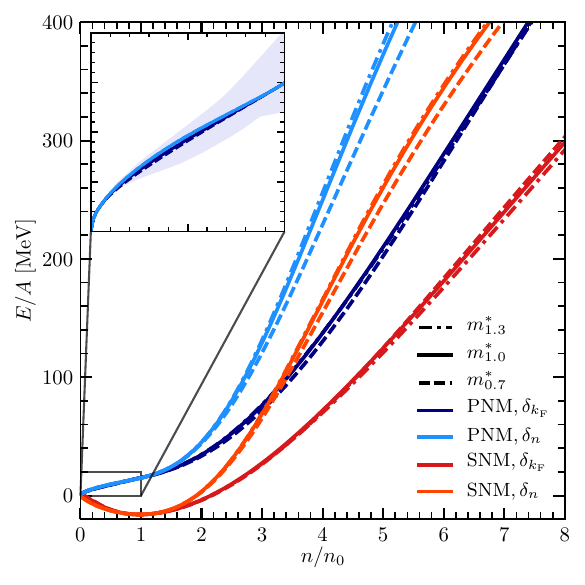}
    \caption{Results for the energy per particle of PNM (blue) and SNM (red) as a function of density for the three effective mass scenarios of Fig.~\ref{fig:effmass_functional}. All depicted EOS employ the smallest $d$ available for the respective choice of $\delta_i$. The nuclear matter properties and high-density input are fixed, see text for details. The light blue band in the inset corresponds to the combined chiral EFT results from Fig.~\ref{fig:energy_constraints}.}
    \label{fig:energy_effmass_variations}
\end{figure}

\begin{figure} 
    \centering
    \includegraphics[width=0.95\columnwidth]{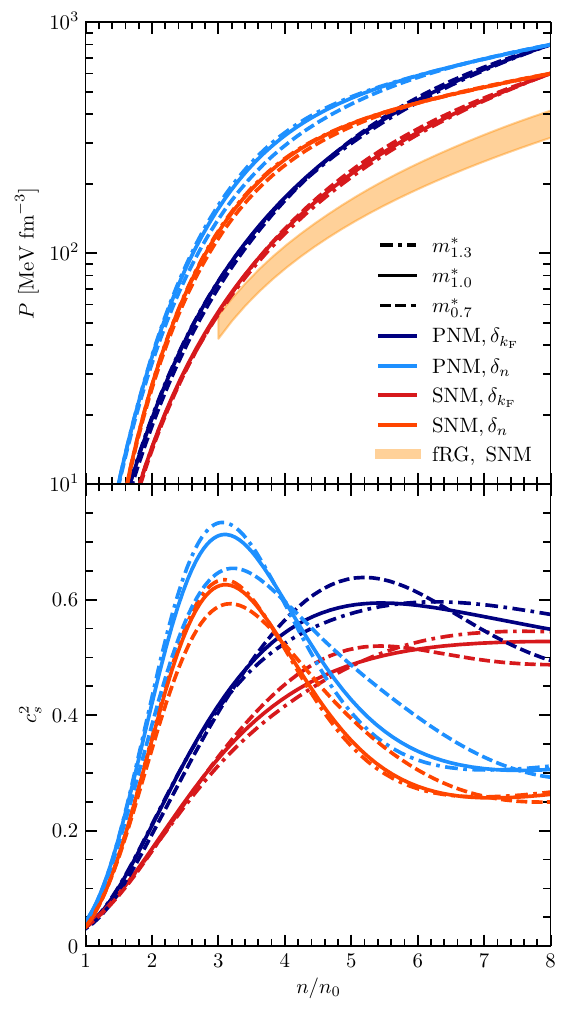}
    \caption{Analog of Fig.~\ref{fig:energy_effmass_variations} for the pressure (upper panel) and speed of sound (lower panel) of PNM (blue) and SNM (red) as a function of density. The orange band corresponds to the fRG SNM result from Ref.~\cite{Leon19fRGeos}.}
    \label{fig:pressure_cs2_effmass_variations}
\end{figure}

As discussed in Sec.~\ref{sec32}, we employ three different parametrizations of the nucleon effective mass, $m^*_{0.7}$, $m^*_{1.0}$, and $m^*_{1.3}$, where the subscript denotes the respective high-density limit, see Fig.~\ref{fig:effmass_functional}. Here, in Figs.~\ref{fig:energy_effmass_variations} and \ref{fig:pressure_cs2_effmass_variations} we examine the impact of the high-density behavior of the effective mass on the zero-temperature EOS. The impact on thermal effects is studied in Sec.~\ref{sec52}.

In Figs.~\ref{fig:energy_effmass_variations} and \ref{fig:pressure_cs2_effmass_variations}, the nuclear matter properties are fixed as
\begin{align}
(K,\ns,B)&= (K_\mathrm{max},(\ns,B)_\mathrm{min}),\\
(\SEv,L)/\MeV&=(30,35).
\end{align}
The high-density input is fixed as specified in Fig.~\ref{fig:pressure_cs2_nucmatt_variations}. We see that the overall influence of the effective mass on the energy and pressure at zero temperature is comparatively small by construction. 
The speed of sound, as a quantity that is not directly constrained by the fit, is more sensitive to variations of the effective mass. 
The observed behavior depends on the choice of $\delta_{i}$, and in each case the results for $m^*_{1.0}$ and $m^*_{1.3}$ are more similar compared to $m^*_{0.7}$. In the case of $\delta_{n}$ the maximum of $c_s^2$ increases with the high-density limit of the effective mass (i.e., the EOS becomes stiffer), while for $\delta_{\kf}$ the speed of sound peak occurs at smaller densities for $m^*_{0.7}$.

\subsection{High-density variations}\label{sec43}

With a careful implementation of the nuclear physics constraints at hand,
see Sec.~\ref{sec41}, the objective now is to have the EOS functional reproduce the high-density constraints from neutron star observations. That is, the goal is to cover the band for neutron star matter obtained by Raaijmakers \textit{et al.}~\cite{Raai19GWNICER} and have EOS that are consistent with the mass measurements of Antoniadis {\it et al}.~\cite{Anto13PSRM201} and the $2\sigma$ confidence interval of the $2.14~M_\odot$ measurement by Cromartie \textit{et al.}~\cite{Crom19massiveNS}, see Sec.~\ref{sec22}. For this, we span a grid of fit points for the pressure of SNM and PNM at $n=1.28\fmiq\approx 8\ns$. We fit the pressure of SNM to values $\{300, 400, 500, 600, 700, 800, 900\} \MeV\fmiq$ and the pressure difference between PNM and SNM to $\{50, 100, 150, 200, 250, 300, 350, 400\}\MeV\fmiq$, so the pressure of PNM ranges from 350 to 1300$\MeV\fmiq$. This results in 56 high-density fit combinations for each low-density and effective mass input. From these, we exclude all EOSs that, after including $\beta$-equilibrium and electrons, are not consistent with the Raaijmakers \textit{et al.}~bands in Fig.~\ref{fig:pressure_constraints}.

The high-density fRG calculations of SNM by Leonhardt \textit{et al.}~\cite{Leon19fRGeos} (see Sec.~\ref{sec23}) lie on the lower end of the employed fit values for the SNM pressure: $310 \MeV\fmiq \lesssim P_\text{fRG}(8\ns) \lesssim 410 \MeV\fmiq$. A lower SNM pressure implies that the pressure of matter in $\beta$-equilibrium is small as well. More specifically, the proton fraction increases with the SNM-PNM pressure difference, leading to a decrease of the pressure of matter in $\beta$-equilibrium. As a result, enforcing consistency with the fRG results reduces the range for neutron star matter to a great extent, such that the uncertainty band by Raaijmakers \textit{et al.}~\cite{Raai19GWNICER} cannot be fully covered. Therefore, we do not use the fRG band as a strict constraint. The subset of EOSs that are consistent with the fRG calculations is studied further in Sec.~\ref{sec51}.

\begin{figure} 
    \centering
    \includegraphics[width=0.95\columnwidth]{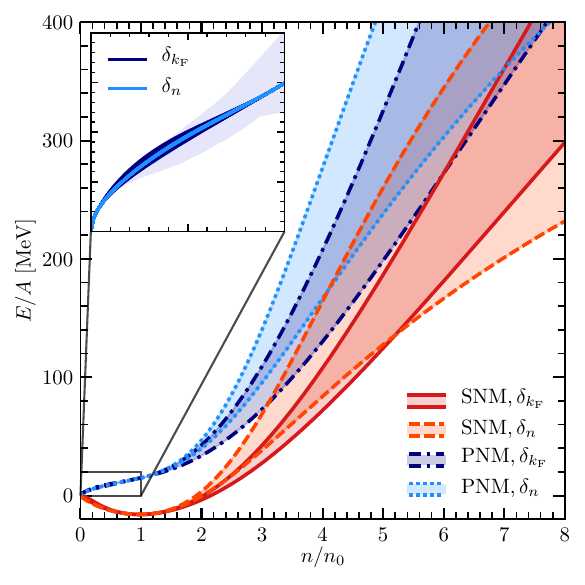}
    \caption{High-density variations for the energy per particle of PNM (blue) and SNM (red) as a function of density, see text for details. The bands for the two different $\delta_i$ combinations include only EOS that are consistent with neutron star constraints. The $d$ parameter, the effective mass, and the nuclear matter properties are kept fixed. The light blue band in the inset corresponds to the combined chiral EFT results from Fig.~\ref{fig:energy_constraints}.}
    \label{fig:energy_highdens_variations}
\end{figure}

\begin{figure} 
    \centering
    \includegraphics[width=0.95\columnwidth]{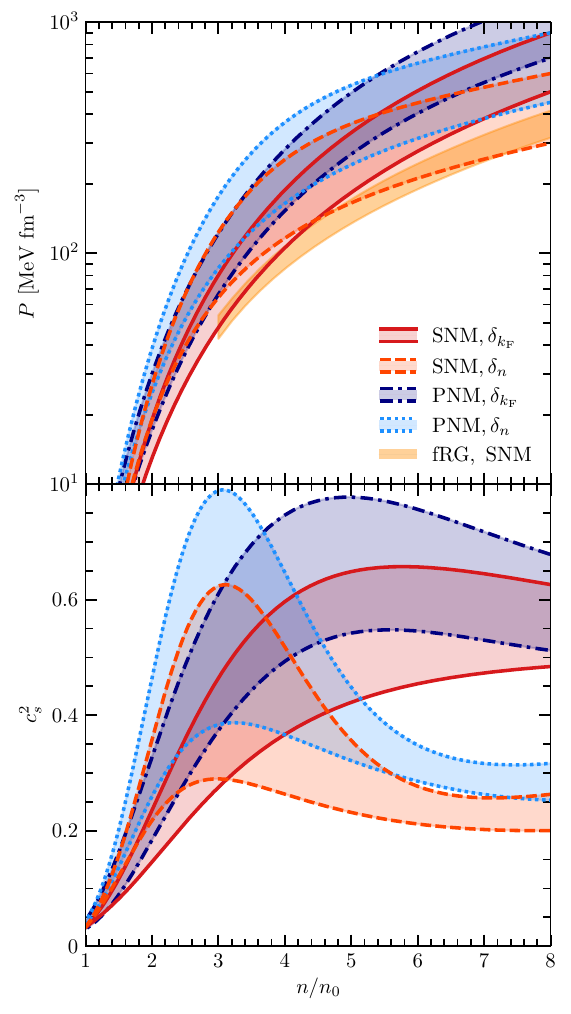}
    \caption{Analog of Fig.~\ref{fig:energy_highdens_variations} for the pressure (upper panel) and speed of sound (lower panel) of PNM (blue) and SNM (red) as a function of density. The orange band corresponds to the fRG SNM results from Ref.~\cite{Leon19fRGeos}.}
    \label{fig:pressure_cs2_highdens_variations}
\end{figure}

As an example, representative high-density variations of the energy per particle of PNM and SNM are shown in Fig.~\ref{fig:energy_highdens_variations}. The corresponding results for the pressure and the speed of sound are displayed in Fig.~\ref{fig:pressure_cs2_highdens_variations}. Here, the nuclear matter properties are set to $(K_\mathrm{max},(\ns,B)_\mathrm{min})$ and $\esym=30\MeV$, $L=35\MeV$. The effective mass is given by $m^*_{1.0}$, and for each $\delta_i$ combination we use the smallest $d$ value from Eq.~\eqref{eq:dsets}. For each $\delta_i$ we keep only the EOSs, which are consistent with the constraints from nuclear physics and neutron star observations. 

As seen in Fig.~\ref{fig:energy_highdens_variations}, for SNM the EOSs with $\delta_{n}$ span a much wider energy band that almost entirely encloses the $\delta_{\kf}$ energy band. At nuclear densities the pressures of both the $\delta_{n}$ and the $\delta_{\kf}$ EOSs lie mostly above the fRG pressure, see the top panel of Fig.~\ref{fig:pressure_cs2_highdens_variations}. At high densities on the other hand the $\delta_{n}$ high-density variations encompass the entire fRG band. In contrast, for $\delta_{\kf}$ the deviations from the fRG band increase with density.

Compared to SNM, the $\delta_{n}$ and $\delta_{\kf}$ energy and pressure bands are of similar size for PNM. Regarding the speed of sound of PNM and SNM, in the bottom panel of Fig.~\ref{fig:pressure_cs2_highdens_variations} one sees that the high-density variations do not lead to significant changes in the systematics for different $(\delta_{i},d)$ choices; see Fig.~\ref{fig:pressure_cs2_functional}. For both $\delta_i$ sets, the EOS with largest stiffness regions involve a PNM speed of sound that at its peak is close to $c_s^2=0.8$. In Fig.~\ref{fig:pressure_cs2_highdens_variations}, for both the PNM and SNM speed of sound the two $\delta_i$ sets give nonoverlapping results at high densities. However, one needs to keep in mind the displayed results are for one particular choice of the $d$ parameter, the nuclear matter properties and the effective mass. Varying these reduces the area that is not covered with the specific input used. Plots that involve the full range of the considered parameter variations are shown in Sec.~\ref{ch5}.

\section{Astrophysical equation of state}\label{ch5}

Here, in Sec.~\ref{sec51} we examine our results for cold matter in $\beta$-equilibrium and study neutron star properties such as the $M\mbox{-}R$ relation and the electron fraction. We take into account the full set of parameter variations of the EOS functional, as discussed in Sec.~\ref{ch4}. Thermal effects, which are crucial for applications in CCSN and NSM simulations, are the subject of Sec.~\ref{sec52}.

\subsection{Neutron star properties}\label{sec51}

The density of electrons and muons in neutron star matter is equal to the proton density because of local charge neutrality. For simplicity, we neglect muons as this causes only a very small change in the neutron star EOS. The proton fraction $x$ at a given baryon density $n$ is fixed by the requirement of $\beta$-equilibrium, i.e., 
\begin{align}
\mu\n(n,x) &=\mu\p(n,x)+\mu_\mathrm{e}(n_\mathrm{e}=xn),
\end{align}
where $\mu_{\text{n},\text{p},\text{e}}$ is the chemical potential of the respective particle species. Electrons can be modeled as an ultrarelativistic degenerate Fermi gas, so the electron pressure is $P_\mathrm{e}=E_\mathrm{e}/(3V)$, with the electron energy density given by $E_\mathrm{e}/V=(3\pi^2 n_\mathrm{e})^{4/3}/(4\pi^2)$. The electron chemical potential reads $\mu_\mathrm{e}=(3\pi^2n_\mathrm{e})^{1/3}$ and the chemical potentials of neutrons and protons are  $\mu_{\text{n},\text{p}}=\partial_{n_{\text{n},\text{p}}} E(n,x)/V + m_{\text{n},\text{p}}$. 

\begin{figure}
    \centering
    \includegraphics[width=0.95\columnwidth]{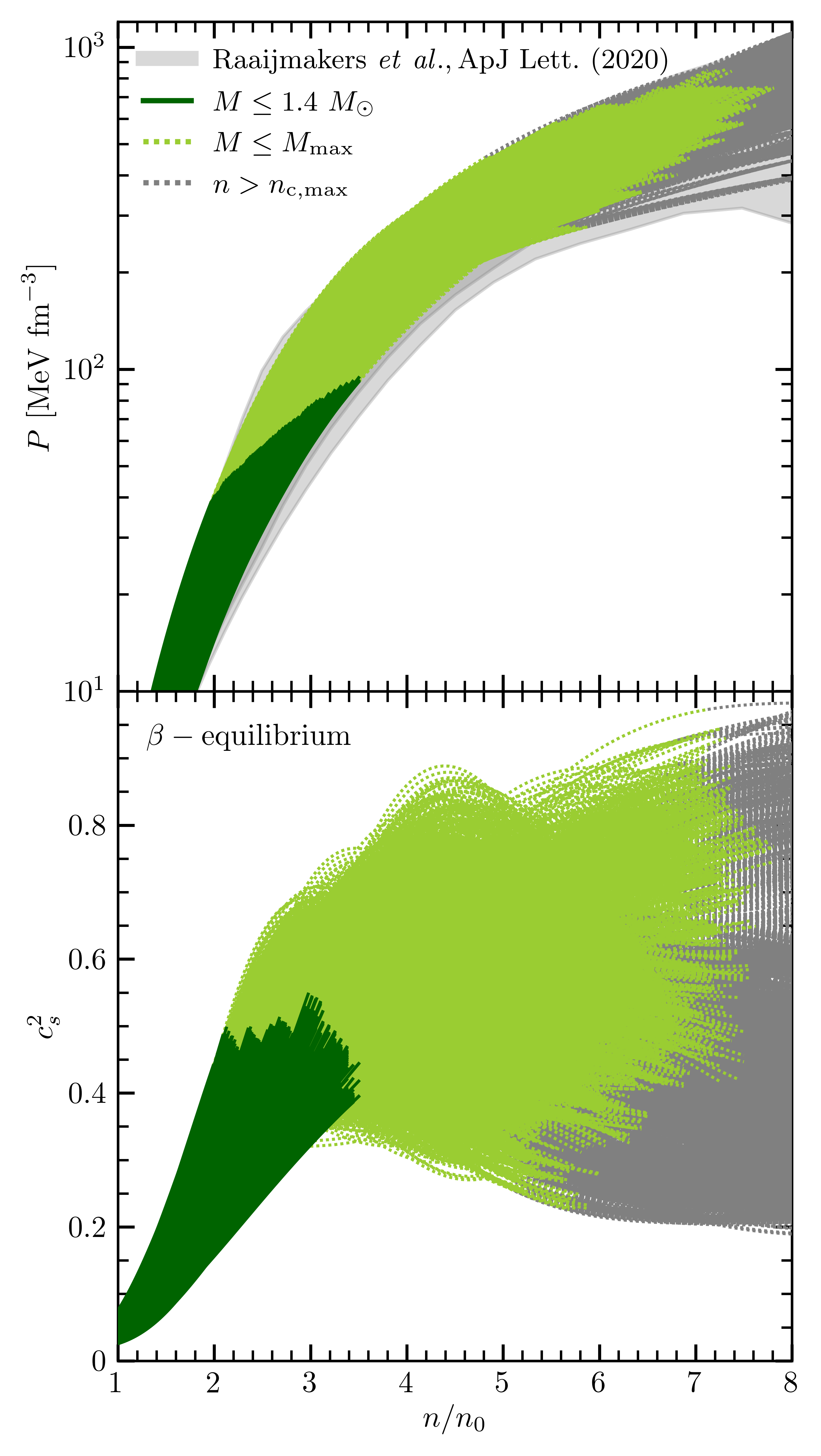}
    \caption{Results for the pressure (upper panel) and the speed of sound (lower panel) of matter in $\beta$-equilibrium as a function of density. 
    As described in the text, we show all EOSs that fulfill the constraints from nuclear physics and neutron star observations.
    The color coding indicates the mass of the corresponding neutron star, where dark green corresponds to masses up to $1.4\ M_\odot$ and light green to higher masses up to the respective maximum mass $M_\mathrm{max}$. Gray lines correspond to the continuation of the EOSs to densities above the central densities $n_\mathrm{c,max}$ of the respective heaviest neutron star. The light-gray band depicts the $95\%$ credible region of the neutron star constraints from Raaijmakers \textit{et al.}~\cite{Raai19GWNICER}.}
    \label{fig:pressure_cs2_beta}
\end{figure}

\begin{figure*}
    \centering
    \includegraphics[width=1.95\columnwidth]{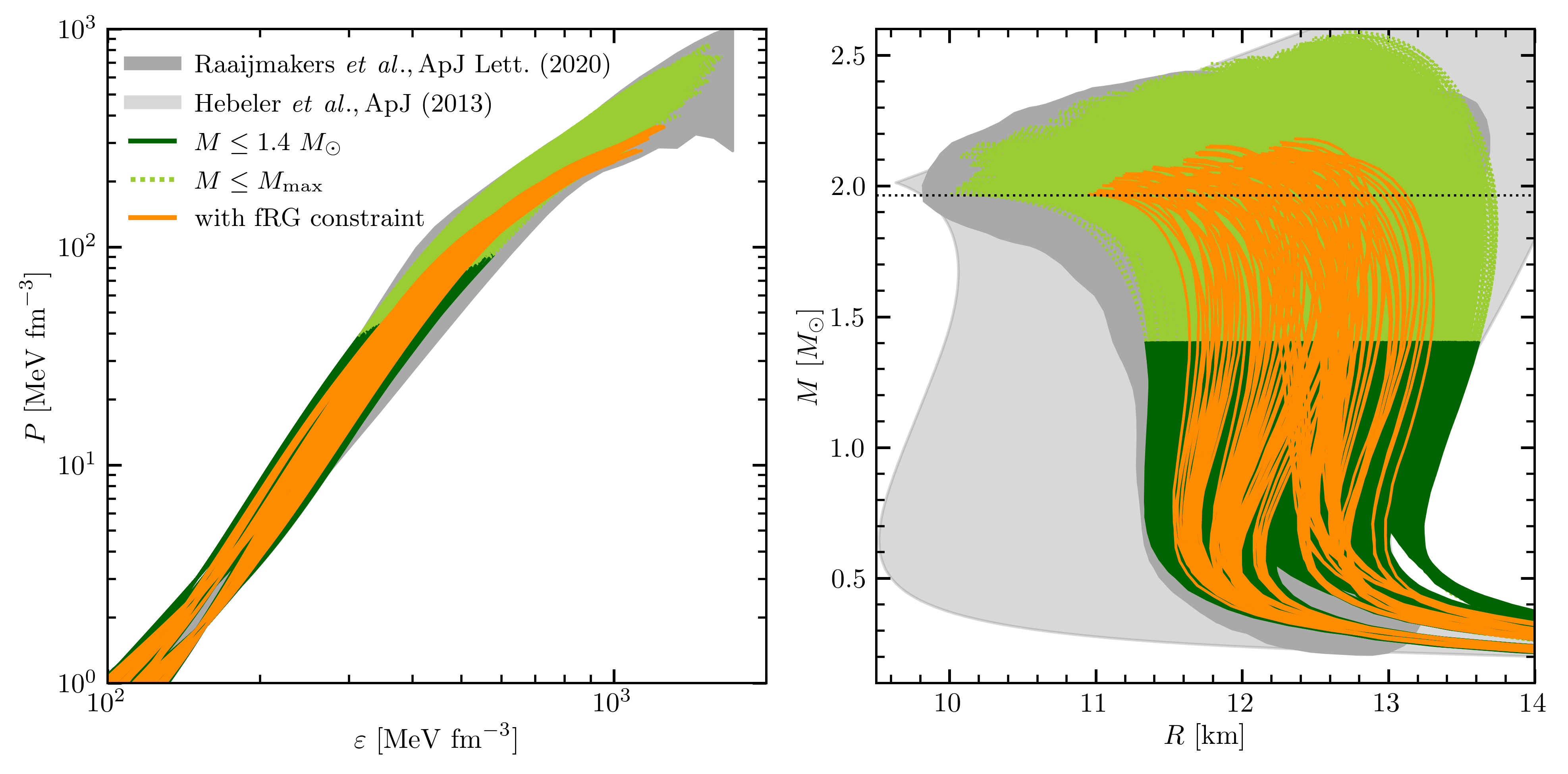}
    \caption{Analog of Fig.~\ref{fig:pressure_cs2_beta} for the pressure-energy density relation (left) and the mass-radius relation (right) of cold neutron stars. The orange lines correspond to EOSs that are consistent with the fRG SNM results from~\cite{Leon19fRGeos}, see text for details. The gray band depicts the $95\%$ credible region of the neutron star constraints from Raaijmakers \textit{et al.}~\cite{Raai19GWNICER}. For comparison, in the $M\mbox{-}R$ plot we show also the uncertainty band obtained by Hebeler \textit{et al.}~\cite{Hebe13ApJ} using piecewise polytrope extensions to high densities (light gray band).}
    \label{fig:EOS_MR}
\end{figure*}

With the variations of the EOS input and the choices for the functional parameters $\delta_i$ and $d$ of Eqs.~\eqref{eq:deltasets} and~\eqref{eq:dsets} in place, we perform all possible fit combinations to obtain bands for the EOS of matter in $\beta$-equilibrium. The neutron star mass-radius $(M\mbox{-}R)$ relation is then obtained by solving the Tolman-Oppenheimer-Volkoff equations~\cite{Tolm39tov,Oppe39tov}. To this end, we implement the Baym, Pethick, Sutherland (BPS) crust from Ref.~\cite{Baym71BPS} for densities below $n_\mathrm{crust}=0.08\fmiq$, as in Ref.~\cite{Raai19GWNICER}. As discussed in the two previous sections, for each of the $\delta_i$ sets we consider four $d$ values, 12 variations of nuclear matter properties, three effective mass scenarios, and 56 high-density fits for the pressure of SNM and PNM, resulting in a set of 16128 EOS. Among these, we keep only EOSs that
\begin{itemize}
    \item are consistent with the theoretical PNM uncertainty band and the unitary gas bound for the energy per particle up to $0.2\fmiq$,
    \item provide masses of neutron stars of at least $1.965\ M_\odot$ (combined lower bound of the measurements from Ref.~\cite{Anto13PSRM201} and the $2\sigma$ interval of Ref.~\cite{Crom19massiveNS}), 
    \item and lie within the $95\%$ credible regions based on the joint analysis of GW170817 and NICER from Raaijmakers \textit{et al.}~\cite{Raai19GWNICER}.
\end{itemize}

The results for the pressure and the speed of sound of neutron star matter are shown in Fig.~\ref{fig:pressure_cs2_beta}. We see that our EOS functional covers almost the entire band for the pressure by Raaijmakers~\textit{et al.}~\cite{Raai19GWNICER}. At the high-pressure boundary the agreement is very close, but some of the softer EOSs within the Raaijmakers \textit{et al.} band are not reproduced. This feature can be largely attributed to the fact that we use a strict lower bound for the minimal value of $M_\text{max}$ whereas Raaijmakers \textit{et al.} have modeled the mass likelihood function for the $2.14\Msun$ pulsar~\cite{Crom19massiveNS}.

In Fig.~\ref{fig:pressure_cs2_beta}, the parts of the different EOSs that correspond to neutron stars with masses below the canonical $1.4\ M_\odot$ are highlighted in dark green. Their continuation up to the respective maximum mass $M_\mathrm{max}$ is colored in light green. The central density $n_{1.4}$ for a neutron star with $1.4\ M_\odot$ lies approximately within $2$--$3.5\,\ns$. The smallest and largest $n_\mathrm{max}$ are roughly $4.5\,\ns$ and $7.5\,\ns$, respectively, where one particular EOS reaches $n_\mathrm{max}\approx 7.9\,\ns$, very similar to Ref.~\cite{Hebe13ApJ}. Up to $n_{1.4}$, the speed of sound is relatively strongly constrained, but at higher densities a large variety of speed of sound curves is present that covers a range from $c_s^2 \approx 0.2$ to almost the speed of light.

Next, in Fig.~\ref{fig:EOS_MR} we show the corresponding results for the pressure-energy density relation and the neutron star $M\mbox{-}R$ diagram. Regarding the $M\mbox{-}R$ relation, compared to our results the band of Raaijmakers {\it et al.}~\cite{Raai19GWNICER} includes $M\gtrsim 1.5\ M_\odot$ neutron stars  with slightly smaller radii. This is a direct consequence of the softer EOSs included there, as discussed above. For a $1.4 \Msun$ neutron star we find a radius range of $R_{1.4}=11.1$--$13.6\km$, similar to Ref.~\cite{Raai19GWNICER}. Interestingly, compared to Ref.~\cite{Raai19GWNICER} our EOS functional gives lower-mass neutron stars with slightly larger radii as well as larger maximum masses for neutron stars with $12\lesssim R/\km\lesssim 13$. Further, in Fig.~\ref{fig:EOS_MR} we also show the mass-radius band obtained by Hebeler \textit{et al.}~\cite{Hebe13ApJ} using polytropic extrapolations of chiral EFT results. Compared to the other bands, the band by Hebeler \textit{et al.}~\cite{Hebe13ApJ} allows for neutron stars with smaller radii and larger maximum masses. This is mainly because it shows the entire region ($100\%$ instead of $95\%$ credible) compatible with the maximum mass constraint.

The density exponents of the functional $\delta_i$ mostly influence the stiffness of the EOS. In particular, softer EOSs corresponding to neutron stars with smaller radii mainly involve $\delta_{\kf}$, while $\delta_\mathrm{n}$ yields stiffer EOSs and larger radii. Moreover, the back-bending of the $M\mbox{-}R$ lines at $M\approx 0.5\ M_\odot$ is more pronounced for EOSs that use $\delta_\mathrm{n}$. 

The EOSs for which SNM is consistent with the fRG band by Leonhardt \text{et al.}~\cite{Leon19fRGeos} are highlighted in orange in Fig.~\ref{fig:EOS_MR}. More specifically, for the orange EOSs the pressure of SNM starting at $5\ns$ deviates from the fRG band by at most $10\%$. Compared to the full band, the fRG-consistent neutron star EOSs have lower pressures at energy densities $\varepsilon\gtrsim 800\MeV\fmiq$, which is a consequence of the relatively low SNM pressures obtained by the fRG calculation. This translates into comparatively larger neutron star radii and smaller maximum masses. Nevertheless, the fRG-consistent EOSs cover a broad range of the pressure-energy density and $M\mbox{-}R$ bands of Raaijmakers {\it et al.}~\cite{Raai19GWNICER}. Overall, the fRG results for SNM provide viable additional constraints for astrophysical EOS constructions, and incorporating them leads to a considerable narrowing of the uncertainty band for the EOS and the $M\mbox{-}R$ relation. In the future, improved fRG calculations will enable further advancements along these lines.

\begin{figure}
    \centering
    \includegraphics[width=0.95\columnwidth]{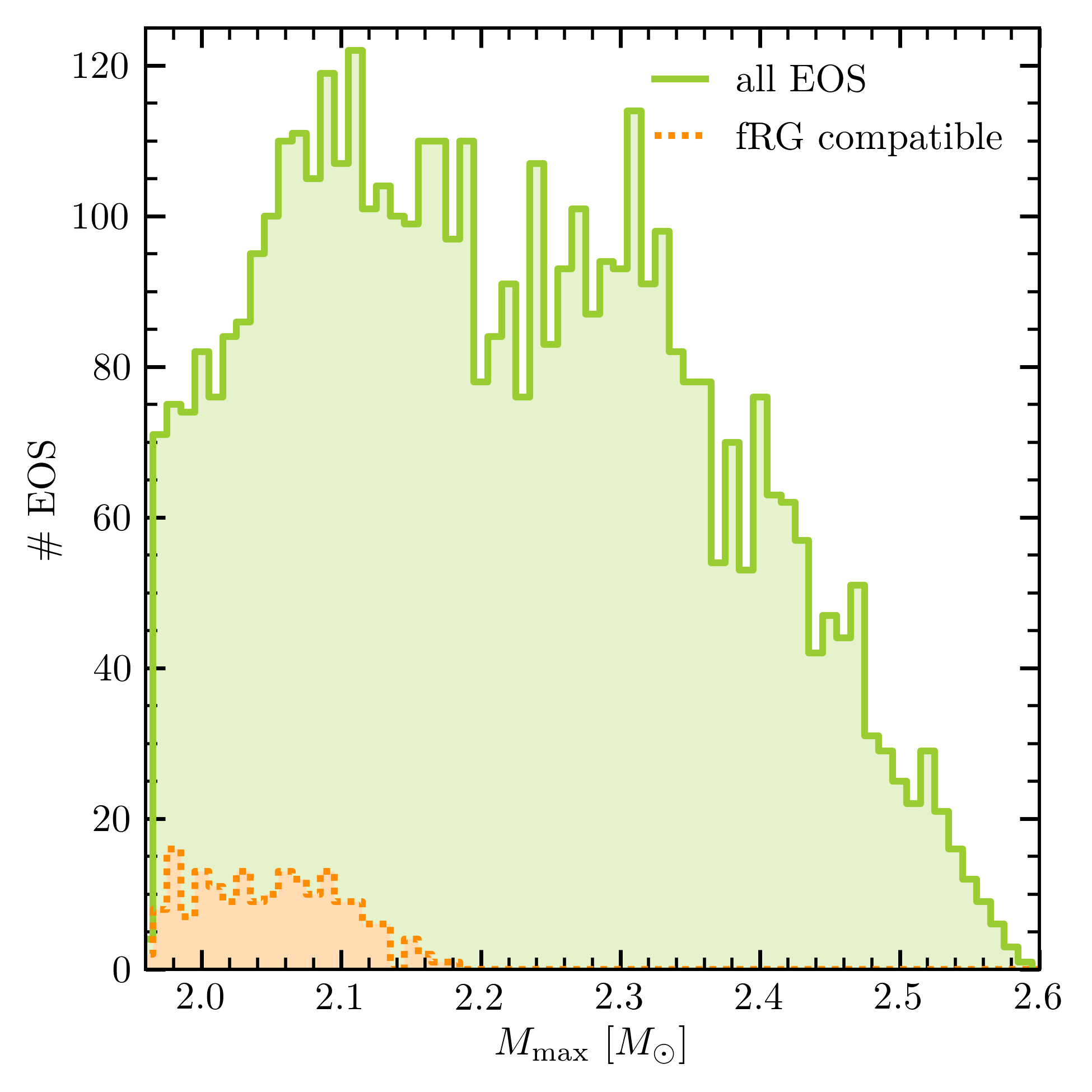}
    \caption{Number of EOSs per maximum mass $M_\mathrm{max}$ corresponding to the mass-radius relation of Fig.~\ref{fig:EOS_MR}. The orange dotted line corresponds to EOSs that are consistent with the fRG SNM results from Ref.~\cite{Leon19fRGeos}.}
    \label{fig:Mmax}
\end{figure}

Our results for the neutron star maximum mass are examined further in Fig.~\ref{fig:Mmax} where we plot the number of EOSs that yield a given value of $M_\text{max}$. The distribution in Fig.~\ref{fig:Mmax} shows a broad peak, which falls off steeply for $M_\text{max}\gtrsim 2.35 \Msun$, reaching zero at $M_\text{max}\approx 2.6 \Msun$. Our largest maximum masses are only slightly above the model-dependent bound $M_\text{max}\lesssim 2.3$--$2.4\Msun$ inferred from GW170817~\cite{Marg17Mmax,Shib17GWmodel, Ruiz18Mmax, Rezz18Mmax, Shib19Mmax, Ai20Mmax, Abbo20modelcompare}. Again, in Fig.~\ref{fig:Mmax} the EOSs that are consistent with the fRG band are highlighted in orange. As discussed above, the fRG constraint implies softer EOSs and thus leads to smaller values of the maximum mass with $M_\mathrm{max}\lesssim 2.18\Msun$.

\begin{figure}
    \centering
    \includegraphics[width=0.95\columnwidth]{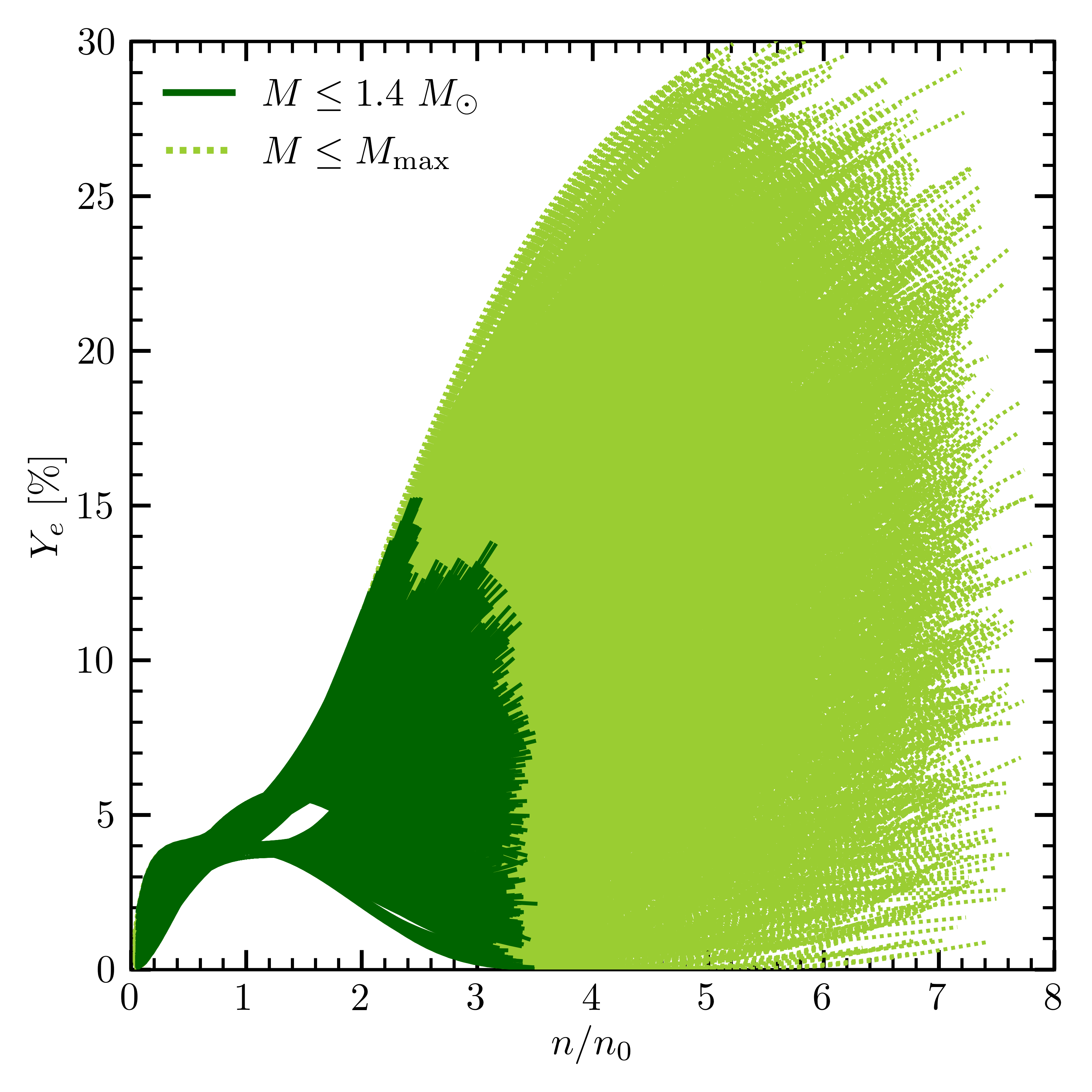}
    \caption{Analog of Fig.~\ref{fig:pressure_cs2_beta} for the electron fraction $\Ye$ as a function of density.}
    \label{fig:Ye}
\end{figure}

Finally, in Fig.~\ref{fig:Ye} we show the electron fraction $\Ye(n)$ in neutron stars as obtained from the different EOS. The density dependence of the electron fraction is strongly related to that of the symmetry energy $\SE(n)$. Similar to the results for the speed of sound, our $\Ye(n)$ band is fairly narrow up to around saturation density where the electron fraction is given by $\Ye(\ns)\approx (4\esym)^3/3\pi^2\ns\approx 0.035$--$0.055$~\cite{Hebe10PRL}. Above $(2$--$3)\ns$, the $\Ye(n)$ band widens considerably, with different EOS given electron fractions between $0$ to about $30\%$ in the core of heavy neutron stars. This reflects the difference in energy of PNM to SNM, as a larger symmetry energy implies a higher electron fraction.

\subsection{Thermal effects}\label{sec52}

Some EOSs used in NSM simulations start from a cold neutron star EOS and add a thermal part that is parametrized in terms of a density-independent thermal index $\Gth=\text{const.}$, e.g., $\Gth=1$--$2$.  The validity of this approximation was studied by Bauswein \textit{et al.} in Ref.~\cite{Baus10thermal} who concluded that a consistent treatment of thermal effects beyond the $\Gth=\text{const.}$ approximation is important. As discussed in Sec.~\ref{sec21a}, the thermal contribution to the EOS is governed to a large extent by the nucleon effective mass. Full EOS tables for astrophysical simulations mostly apply a mean-field effective mass that monotonically decreases with density. This, however, is not consistent with microscopic nuclear-matter calculations~\cite{Carb19thermalEOS,Kell20finiteT}, which show that interaction effects beyond the mean-field approximation are significant. 
As discussed in Sec.~\ref{sec31}, our EOS functional incorporates such microscopic effective-mass results explicitly.

The temperature dependence of our EOS functional is described solely by the kinetic term, which is modeled as a noninteracting nucleon gas with neutron and proton effective mass $m^*_\text{n}(n,x)$ and $m^*_\text{p}(n,x)$, see Sec.~\ref{sec32}. From this one arrives at the following equation for the thermal index $\Gth$ of isospin-asymmetric nuclear matter (ANM) with proton fraction $x$: 
\begin{align}\label{eq:Gth_m*ANM}
    \Gth(n,x,T) 
    &= \frac{5}{3}-
    n\frac{\sum_{t}\frac{\varepsilon_{t,\text{th}}(n,x,T)}{m^*_t(n,x)}\frac{\partial m^*_t(n,x)}{\partial n}}
    {\sum_{t} \varepsilon_{t,\text{th}}(n,x,T)},
\end{align}
where $\varepsilon_{t,\text{th}}(n,x,T)=\varepsilon_{t}(n,x,T)-\varepsilon_{t}(n,x,0)$ is the thermal part of the kinetic energy density of neutrons and protons, respectively. In the context of our EOS functional, Eq.~\eqref{eq:Gth_m*ANM} is an exact representation of $\Gth(n,x,T)$, i.e., it is equivalent to Eq.~\eqref{GammathDef}. The $T$ dependence of $\Gth(n,x,T)$ vanishes for $x=0$ and $x=1/2$, i.e., for PNM and SNM one obtains the familiar expression given by Eq.~\eqref{GammathM*}. However, for ANM the thermal index is a temperature-dependent quantity. (Equation~\eqref{eq:Gth_m*ANM} would reduce to Eq.~\eqref{GammathM*} also for ANM if the effective masses of protons and neutron were identified, $m^*_\text{n}(n,x)=m^*_\text{p}(n,x)$, but this would be in conflict with nuclear theory results, see Sec.~\ref{sec31}.)

Consequently, we explore whether the $T$ dependence of $\Gth$ for ANM is a significant effect, and, if the $T$ dependence is small, what is the appropriate temperature-independent approximative expression for the thermal index. For a classical free nucleon gas with
$m^*_\text{n}(n,x)$ and $m^*_\text{p}(n,x)$ one obtains by substituting $3Tn_t/2$ for $\varepsilon_{t,\text{th}}(n,x,T)$ in Eq.~\eqref{eq:Gth_m*ANM} the expression
\begin{align}\label{eq:Gth_m*ANMclass}
    \Gamma_\text{th,classical}(n,x)
    &= \frac{5}{3}-
    \sum\limits_{t}
    \frac{n_{t}(n,x)}{m^*_t(n,x)}\frac{\partial m^*_t(n,x)}{\partial n},
\end{align}
with $n_\text{n}=(1-x)n$ and $n_\text{p}=xn$. Note that for $x=0$ and $x=1/2$, respectively, both Eq.~\eqref{eq:Gth_m*ANM} and Eq.~\eqref{eq:Gth_m*ANMclass} yield the correct equation for PNM and SNM, Eq.~\eqref{GammathM*}.

Comparing the results obtained from Eqs.~\eqref{eq:Gth_m*ANM} and~\eqref{eq:Gth_m*ANMclass} one finds that the classical expression provides a very good approximation, with relative errors well below the $1\%$ level (except at very low $T\lesssim 1$ MeV). For example, averaging over densities $n/\ns\in[0,8]$ and the results obtained from the three effective mass scenarios ($m^*_{0.7}$, $m^*_{1.0}$, and $m^*_{1.3}$), the mean relative errors at $T=10$ MeV are $(0.18\%,0.52\%,0.63\%)$ for $x=(0.1,0.2,0.3)$. At higher temperatures one is closer to the classical limit, so the errors decrease with $T$; at $T=1$ MeV and $T=50$ MeV they are  $(0.22\%,0.61\%,0.85\%)$ and $(0.09\%,0.16\%,0.24\%)$, respectively, for $x=(0.1,0.2,0.3)$. 
For each $T$ and $x$ and each effective mass scenario the deviations first increase with density up to $n/\ns\approx 5$--$7$, and then decrease again (since the high-density limit of the effective mass is $x$ independent in our approach). Overall, we conclude that $\Gamma_\text{th,classical}(n,x)$ provides a very good representation of the temperature dependence of the EOS of ANM.

\begin{figure}
    \centering
    \includegraphics[width=0.95\columnwidth]{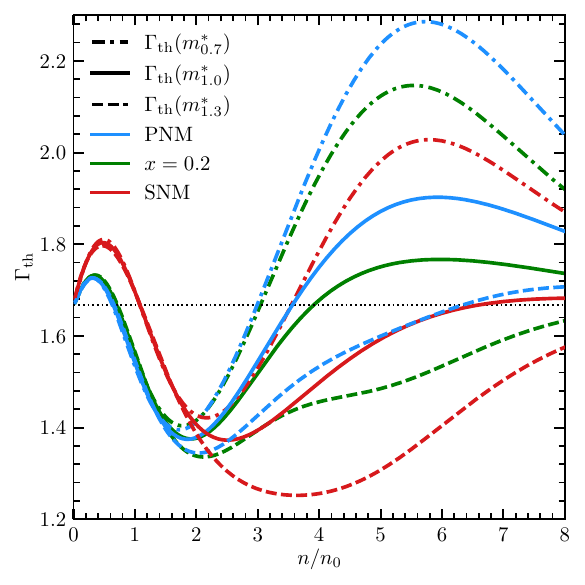}
    \caption{Results for the thermal index of PNM (blue),  SNM (red), and ANM with $x=0.2$ (green) as a function of density.     
    The different line types correspond to the three effective mass scenarios, see Sec.~\ref{sec31}. 
    The horizontal gray dotted line corresponds to the thermal index $\Gth=5/3$ of a free or unitary Fermi gas.}
    \label{fig:Gamma_th}
\end{figure}

Our results for the thermal index $\Gth$ of PNM, SNM, and ANM with $x=0.2$ are shown in Fig.~\ref{fig:Gamma_th}, where for ANM we show the results obtained from the classical approximation, Eq.~\eqref{eq:Gth_m*ANMclass}. The density behavior of $\Gth$ is then for each $x$ determined entirely by that of $m^*_\text{n}(n,x)$ and $m^*_\text{p}(n,x)$. For PNM and SNM, an increasing (decreasing) effective mass implies that $\Gth$ is below (above) the free or unitary Fermi gas value $\Gth=5/3$, and the thermal index of PNM is larger than that of SNM. For each $x$, at low densities $\Gth$ first increases with $n$ and then decreases again such that $\Gth=5/3$ is reached at $n\approx \ns$, corresponding to the minimum of $m^*_t(n,x)$ at around saturation density (see Sec.~\ref{sec31}). The high-density behavior is fixed by the respective effective mass scenario, where the largest deviations from $\Gth=5/3$ occur for $m^*_{0.7}$ at $n\approx 5.8\ns$ (the $n\rightarrow 0$ limit is $\Gth\rightarrow 5/3$ by construction). The smallest values, e.g., $\Gth\approx 1.25$ at $n\approx 3.7\ns$ for SNM, are obtained for $m^*_{1.3}$.

The detailed description of thermal effects within our EOS functional may have interesting effects in astrophysical applications. In particular, the proto-neutron star contraction in CCSN simulations is largely governed by the $T$ dependence of the EOS~\cite{Yasi20EOSeffects,Schn17LSEOS}. Lower effective masses lead to larger thermal contributions and thus to a larger PNS radius. A faster contraction increases the temperature at the surface of the PNS. As a consequence, neutrinos emitted from the PNS have larger energies, which aids the shock evolution towards a faster explosion. All effective mass scenarios result in a thermal index that is mostly well below $5/3$ at $(1$--$2)n_0$, which may lead to a faster PNS contraction and explosion compared to commonly used astrophysical EOS such as the Lattimer-Swesty or Shen EOS~\cite{Latt91LSEOS,Latt85LLPR,Shen98eos}. The very high-density regime of the EOS is more important in NSM than in CCSN applications. Investigating the effects of our different high-density effective mass scenarios in NSM simulations may be an interesting subject for future research.

\section{Summary and outlook}\label{ch6}

In this paper, we have developed a new EOS functional for application in CCSN and NSM simulations. The EOS functional is constrained by various chiral-EFT based calculations of neutron matter and nuclear matter, by astrophysical observations, and (to a lesser extent) by results from a recent QCD-based fRG study of high-density matter. In particular, the EOSs obtained from our functional are consistent with recent mass measurements of heavy neutron stars from Refs.~\cite{Anto13PSRM201,Crom19massiveNS} and the joint analysis of observational data from GW170817 and NICER from Raaijmakers {\it et al.}~\cite{Raai19GWNICER}. 

Using as input the recent microscopic nuclear-matter results from Ref.~\cite{Carb19thermalEOS}, we have modeled the kinetic part of the EOS as a noninteracting nucleon gas with density-dependent effective mass. The careful implementation of microscopic results for the nucleon effective mass in our EOS functional is an essential novelty compared to previous constructions of astrophysical EOS~\cite{Latt91LSEOS,Shen98eos,Shen11eos3,Hemp12EOSccsn,Stei13EOSccsn,Schn17LSEOS}, in particular, the commonly used Lattimer-Swesty EOS~\cite{Latt91LSEOS} and Shen EOS~\cite{Shen98eos}.

The effective mass is a key quantity that determines the temperature dependence of the EOS. It was demonstrated in various studies~\cite{Yasi20EOSeffects,Schn19EOSeffects,Baus10thermal} that a thorough description of the effective mass is crucial for CCSN and NSM simulations. In this work, we have studied different high-density extrapolations of the microscopic nuclear-density results for the effective mass. It will be interesting to investigate the impact of different effective mass scenarios on thermal effects in astrophysical applications. 

\begin{figure}
    \centering
    \includegraphics[width=0.95\columnwidth]{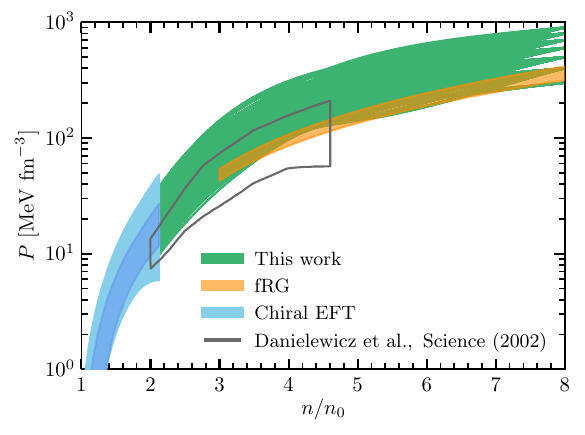}
    \caption{Our results for the pressure of SNM as a function of density (green band) in comparison to constraints from chiral EFT (blue bands) and the fRG study (orange band) from Ref.~\cite{Leon19fRGeos}, see Sec.~\ref{sec23}. In addition, we show constraints extracted from heavy-ion collisions (gray outline) from Ref.~\cite{Dani02Esymm}.}
    \label{fig:PSNM}
\end{figure}

The description of the interaction part in our approach represents an improvement over traditional EOS functionals~\cite{Latt91LSEOS,Shen98eos,Shen11eos3,Hemp12EOSccsn,Stei13EOSccsn,Schn17LSEOS} as well. That is, we have modeled the interaction part not as a sum of density monomials but as a sum of density-dependent rational functions (the temperature-independence of the interaction part is small~\cite{Kell20finiteT}, and thus was neglected in this work). This ensures that the EOS functional is appropriately stable under variations of the low- and high-density input. We have fitted the parameters of the interaction part to the combination of state-of-the-art neutron matter calculations and observational constraints. From this, we have derived a comprehensive uncertainty band for neutron star matter. Our EOSs predict that the radius of a canonical 1.4 solar mass neutron star lies in the range $R_{1.4}=11.1$--$13.6\km$. The fRG results of Leonhardt \textit{et al.}~\cite{Leon19fRGeos} provide an additional constraint that leads to a significant reduction in the radius uncertainty of neutron stars. As a summary plot of our results for SNM, we show in Fig.~\ref{fig:PSNM} the pressure of SNM from our EOS functional, as well as from the chiral EFT and fRG results from Ref.~\cite{Leon19fRGeos}, and compare these with the constraints extracted from heavy-ion collisions from Ref.~\cite{Dani02Esymm}. The comparison of these results shows that a very promising consistency is emerging between these different dense-matter constraints.

Overall, our EOS functional represents a significant step towards a microscopic description of the complete uncertainty range of the dense matter EOS for astrophysical simulations. Future work will be targeted at further refinements of the EOS functional, including the careful consideration of possible phase-transition effects on the dense matter EOS, and the construction of comprehensive EOS tables for direct use in CCSN and NSM simulations.

\section*{Acknowledgments}

We thank Almudena Arcones, Arianna Carbone, Svenja Greif, Kai Hebeler, Jonas Keller, Yeunhwan Lim, Mirko Pl{\"o}{\ss}er, Geert Raaijmakers, and Ingo Tews for useful discussions. This work was supported by the Deutsche Forschungsgemeinschaft (DFG, German Research Foundation) -- Project-ID 279384907 -- SFB 1245 and benefited from discussions within IReNA, which is supported in part by the National Science Foundation under Grant No.~OISE-1927130.

\bibliography{strongint.bib}
\bibliographystyle{apsrev4-1}

\end{document}